\documentclass{aa}
\bibpunct{(}{)}{;}{a}{}{,} 
\usepackage{graphicx}
\usepackage[varg]{txfonts}
\newcommand{\IRAF}  {{\sc iraf}}

\newcommand{\tim}[1]{\ensuremath{\times 10^{#1}}}
\def\l{\ensuremath{\lambda}}

\newcommand{\Msun}{M$_\odot$}
\newcommand{\kms}{km\,s$^{-1}$}

\newcommand{\ergs}{erg\,s$^{-1}$cm$^{-2}$}

\newcommand{\Halpha} {H$\alpha$}
\newcommand{\Hbeta}  {H$\beta$}
\newcommand{\Hgamma} {H$\gamma$}
\newcommand{\Hdelta} {H$\delta$}

\newcommand{\FeII} {Fe\,{\sc ii}}
\newcommand{\HeI}  {He\,{\sc i}}
\newcommand{\HeII} {He\,{\sc ii}}

\begin{document}

\title{Voracious vortexes in cataclysmic variables:}
\subtitle{A multi-epoch tomographic study of HT Cassiopeia}
\titlerunning{Multi-epoch tomographic study of HT Cas}

\author{V. V. Neustroev\inst{1} \and S. V. Zharikov\inst{2} \and N. V. Borisov\inst{3}}
\institute{Astronomy and Space Physics, PO Box 3000, FIN-90014 University of Oulu, Finland\\
          \email{vitaly@neustroev.net}
          \and
           Instituto de Astronom{\'i}a, Universidad Nacional Aut{\'o}noma de M{\'e}xico,
           Apdo. Postal 877, Ensenada, 22800 Baja California, M{\'e}xico\\
          \and
           Special Astrophysical Observatory of the Russian AS, Nizhnij Arkhyz, Karachaevo-Cherkesia
           369167, Russia
           }
\date{Received April 20, 2015; accepted November 4, 2015}


\abstract
 {We present multi-epoch, time-resolved optical spectroscopic observations of the dwarf nova HT~Cas,
 which were obtained during 1986, 1992, 1995, and 2005 with the aim of studying the properties of emission
 structures in the system. We determined that the accretion disc radius, measured from the double-peaked
 emission-line profiles, is consistently large and lies within the range of 0.45--0.52$a$, where $a$
 is the binary separation. This is close
 to the tidal truncation radius $r_{\rm max}$=0.52$a$. However, this result is not consistent with
 previous radius measurements. An extensive set of Doppler maps  reveals a very complex
 emission structure of the accretion disc. Apart from a ring of disc emission, the tomograms display at
 least three areas of enhanced emission: the hot spot from the area of interaction between the gas stream
 and the disc, which is superposed on the elongated spiral structure, and the extended bright
 region on the leading side of the disc, which is opposite to the location of the hotspot.
 The position of the hotspot in all the emission lines is consistent with the trajectory of the gas stream.
 However, the peaks of emission are located in the range of distances 0.22--0.30$a$, which are much closer
 to the white dwarf than the disc edge. This suggests that the outer disc regions have a very low density,
 allowing the gas stream to flow almost freely before it starts to be seen as an emission source.
 We have found that the extended emission region on the leading side of the disc is always observed at the
 very edge of the large disc. Observations of other cataclysmic variables, which show a similar emission
 structure in their tomograms, confirm this conclusion. We propose that the leading side bright region
 is caused by irradiation of tidally thickened sectors of the outer disc, by the white dwarf and/or hot
 inner disc regions.
}

\keywords{methods: observational -- accretion, accretion discs -- binaries: close --
             novae, cataclysmic variables -- stars:individual: HT~Cas
               }

\maketitle
\section{Introduction}

Cataclysmic variables (CVs) are close interacting binary systems that consist of a white dwarf (WD) as
primary and a low-mass main-sequence star or a brown dwarf as secondary component \citep[and references
therein]{Warner:1995aa}. The Roche-lobe-filling secondary loses matter via the inner Lagrangian point
to the primary. In the absence of a strong magnetic field, the material transferred from the donor star
forms an accretion disc around the WD and gradually spirals down onto its surface where it eventually accretes.

Typical optical spectra of CVs are dominated by emission lines of hydrogen and neutral helium series,
which are formed in the accretion disc. Lines of other species, such as those from singly ionized helium, calcium,
and iron are also often seen \citep{Williams83,Honeycutt87}. The emission lines of CVs with a moderately
high orbital inclination are usually very broad with a full-width velocity over several thousand \kms\
and have a double-peaked profile, which  resuls from the Doppler shift of matter that is rotating in a Keplerian disc
\citep{Smak1969,Smak1981,HorneMarsh86}.

Although the accretion disc is the dominant light source in CVs, many examples show that other emission
components may distort the originally symmetric line profile. It is commonly observed that the intensities
of the red and blue peaks of the double-peaked profiles are variable with the orbital period phase
\citep{GreensteinKraft}. The trailed spectra often show a narrow emission component that moves from
one line hump to the other and back during the course of the orbital period, having the form of an
``S-wave'' \citep{Kraft62}. This S-wave component is usually attributed to a region of high temperature
and luminosity at the outer edge of the accretion disc, which is caused by its interaction with the inflowing
gas stream \citep{Smak1970}. This interpretation is supported by the phasing of the S-wave component, which
crosses from blue-shifted to red-shifted around phase 0.8--0.9, and which corresponds closely to the expected
phasing of this bright area. For the remainder of this paper, we use the slightly outdated term ``the hotspot''
to refer to the area of interaction between the gas stream and the accretion disc to distinguish
the latter from other bright spots on accretion discs.

Soon after, it became apparent that other sources of emission may also be present in an accretion disc. In 1981
\citeauthor{YSS} reported the spectroscopic study of the dwarf nova \object{HT~Cas},
whose spectra showed unusual
behaviour: ``The blue wing is stronger near phase 0.0 and the red one stronger at phase 0.5. This
resembles the variations to be expected from an S-wave, but is 180\degr\ out of phase!''. The
identification of this and other detected emission spots, the phasing of which does not agree with
that expected for the hotspot model, was unclear.

\begin{figure}
\resizebox{\hsize}{!}{\includegraphics{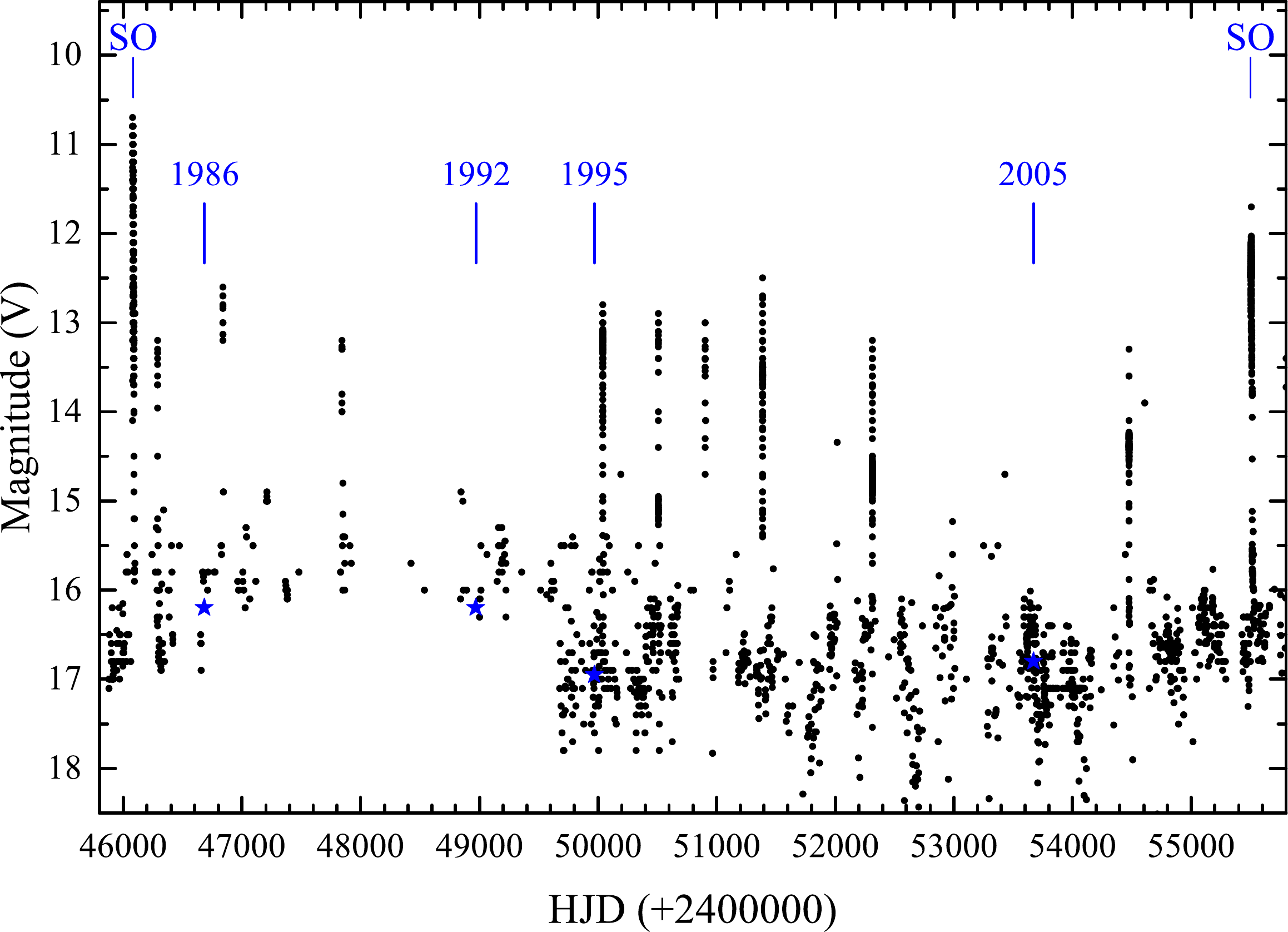}}
\caption{AAVSO light curve of HT Cas between two superoutbursts in 1985 and 2010, marked by ``SO''.
         The blue stars represent averaged magnitudes of the star during our spectroscopic observations.}
\label{Fig:AAVSO}
\end{figure}

The development of Doppler tomography opened up a new regime for the study of accretion structures in
interacting binaries \citep{MarshHorne88}. This technique uses the information encoded in spectral line
profiles that is taken at different orbital phases to calculate a distribution of emission over the binary.
Doppler tomography is now widely used to study interacting binary systems, with the tomograms of  dozens of
CVs have now been  produced. Although a Doppler map is subject to interpretation, since it is created in
velocity space, the predicted location of various binary system components in spatial coordinates can
easily be translated into velocity coordinates and, hence, compared with the map.
Besides the hotspot, this approach helps to identify an irradiated part of secondary stars \citep{MarshHorne90},
spiral structures in accretion discs \citep{Steeghs:1997aa}, accretion flows in polars \citep{PolarDopMap},
and a `reversed bright spot' caused by the deflected gas stream flow that passes above the disc and hits
its back \citep{NeustroevWZ,NeustroevUX}.

Nevertheless, there are observed emission structures that still have no plausible explanation. One of
the most mysterious is a bright spot on the leading side of the disc, opposite to the usual
location of the hotspot. In trailed spectra, this bright spot produces an S-wave which crosses from blue-shifted to
red-shifted around phase 0.5. In Doppler images, it is situated in the bottom-right part of the map.
This place is far from the region of interaction between the stream and the disc particles.
The presence of the leading side bright spot was reported for \object{RR~Pic} \citep{RR_Pic}, \object{WX~Cet}
\citep{Tappert2003}, \object{BZ~UMa} \citep{NeustroevBZ}, \object{VW~Hyi} \citep{VW_Hyi},
\object{1RXS J180834.7+101041} \citep{Yakin}, \object{V406~Vir} \citep{SDSS1238}, \object{EZ~Lyn}
\citep{SDSS0804}, \object{V2051~Oph} \citep{V2051Oph1, V2051Oph2}. Even the tomograms of the famous
\object{IP~Peg} in quiescence occasionally display this kind of bright spot. This list of objects is nowhere near
complete. A rough statistical analysis applied to a sample of 68 CVs with published emission-line profile
studies showed that the presence of the leading side bright spot is not an exception but a frequent
phenomenon \citep{TappertLNP}, the source of which has not yet been found.

For these reasons, we were motivated  to perform a detailed study of the properties of emission structures
in \object{HT~Cas}, seemingly the first CV where a bright spot in the leading side of the accretion disc
has been noticed \citep{YSS}. In this paper we present and discuss the medium-resolution spectroscopic data
obtained during 1986, 1992, 1995, and 2005.

\begin{table*}
\caption{Log of spectroscopic observations of HT Cas}
\label{ObsTab}
\centering
\begin{tabular}{cccccccc}
\hline\hline
 Set        & Date        & Telescope/     &$\lambda$~range& Exp.time & Number   & Duration\\
            &             & instrument     &     (\AA)     & (sec) & of exps. &  (hours)\\
\hline
  Set-1986  & 1986-Sep-08 & 6.0~m / SP-124 &   3600--5500  &  300  &    15    &  1.77 \\
            & 1986-Sep-09 & 6.0~m / SP-124 &   5400--7300  &  360  &    16    &  2.12 \\
  Set-1992  & 1992-Dec-16 & 6.0~m / SP-124 &   3600--5500  &  300  &    14    &  1.41 \\
  Set-1995  & 1995-Sep-09 & 6.0~m / SP-124 &   3860--5800  &  180  &    58    &  3.54 \\
  Set-2005  & 2005-Oct-29 & 2.1~m / B\&Ch  &   4600--6700  &  293  &    40    &  3.54 \\
            & 2005-Oct-31 & 2.1~m / B\&Ch  &   6150--7225  &  293  &    40    &  3.54 \\
\hline
\end{tabular}
\end{table*}

\begin{figure*}[t]
  \includegraphics[width=17cm]{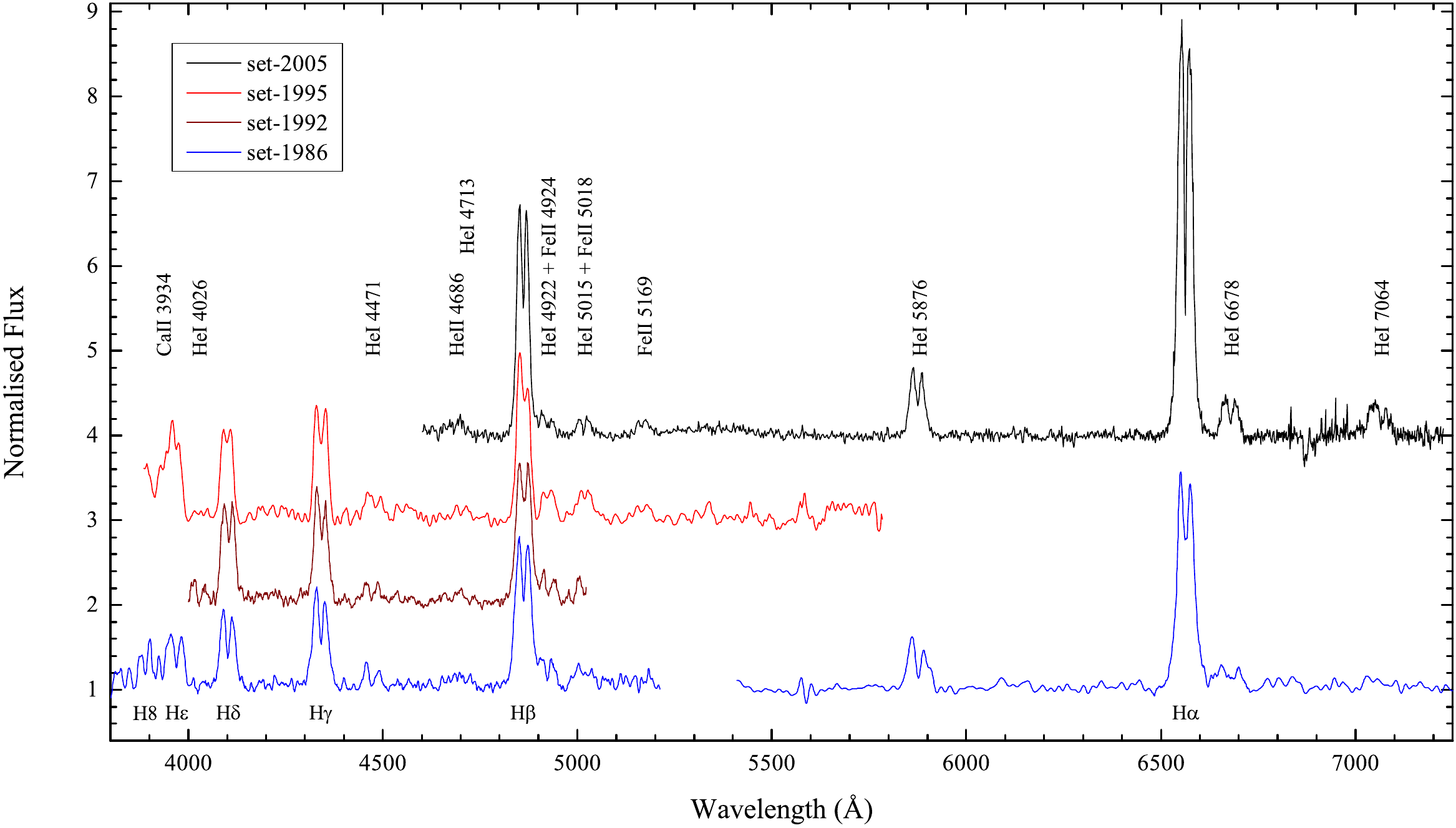}
  \caption{Combined and continuum-normalised out-of-eclipse spectra of HT~Cas for each epoch of observation.
 Spectra shifted vertically to prevent overlap.}
  \label{Fig:aver-spec2}
\end{figure*}

\section{HT Cassiopeiae}
\label{Sec:HT_Cas}

HT~Cas was discovered by \citet{Hoffmeister} and classified as a U~Gem-type star with brightness varying
between 13.0 and 16.5 mag. The eclipses of HT~Cas were discovered by \citet{Bond} and extensively observed by
\citet{Patterson81}, who derived an orbital period of 1.77~h for the system. \cite{Borges2008} report
that the orbital period shows period changes of semi-amplitude $\sim$40~s, which seems to repeat on a
timescale of about 36~yr. The system is characterised by very rare outbursts with mean intervals of
400 days \citep{Wenzel}, which suggests an extremely low mass-transfer rate in the system. Moreover, the
outbursts occur rather irregularly, e.g. no outbursts were observed from 1989 to 1995 \citep{Kato2012}.
In January 1985, \citet{Zhang} observed HT~Cas in a long and bright outburst and detected superhumps in
the light curve of the object, confirming its classification as an SU UMa star. Only two superoutbursts
are documented for HT~Cas; the only other one after 1985 was observed  in 2010 \citep{Kato2012}. The system
is also known to exhibit large-amplitude, long-timescale quiescent light variations, from about 15.9
to 17.7 mag \citep[see also Fig.~\ref{Fig:AAVSO}]{LongTerm}.

Eclipses provide a great opportunity to measure system parameters and to study the accretion disc
structure in detail. \citet{Patterson81} even called HT Cas ``the Rosetta Stone of dwarf novae'', yet
this system might appear to be too atypical for dwarf novae. The light curves and eclipse profiles of
HT~Cas  vary greatly \citep{Patterson81,Zhang,Wood95}. They predominantly show the deep eclipse of the
WD, also visible in X-rays \citep{Wood95}, but the accretion disc component is usually weak and sometimes
absent. Orbital humps -- hotspot modulations -- are rarely seen in quiescence. However, the brightness
of the hotspot increased significantly during the 2010 superoutburst \citep{Bc-akowska:2014aa}.

The system parameters for HT~Cas are usually taken from \citet{Horne91}: $M_1 = 0.61\pm0.04M_\odot$,
$M_2=0.09\pm0.02M_\odot$, $q=0.15\pm0.03$, $i=81.0\pm1.0$\degr. They are based on the derived parameters of
a WD eclipse. Several authors studied the accretion disc in HT~Cas with the multicolour eclipse-mapping
technique. These studies showed that the quiescent disc has a flat brightness temperature profile
(5000--7000~K) \citep{Wood92} and is probably patchy \citep{Vrielmann2002}. Using a similar approach,
\citet{Ultracam} reveal changes in the quiescent accretion disc structure, possibly related to
variations in the mass-transfer rate from the secondary star. There has been controversy in the
literature over the optical thickness of the accretion disc in HT~Cas.
\citet{Zhang} found that the disc is optically thick, \citet{Vrielmann2002} argue that the disc is
moderately optically thin, but becomes optically thick near the WD, while \citet{Wood92} and
\citet{Ultracam} claim that the disc is optically thin in both its inner and outer regions.

The optical spectrum of HT~Cas is typical for a high-inclination dwarf nova, with strong double-peaked
emission lines of hydrogen and neutral helium. First spectroscopic observations were made
by \citet{Rafanelli} who presented a few photographic spectra. Unfortunately, despite a rich history
of photometric investigations, HT~Cas has been almost neglected by optical high- or medium-resolution
spectroscopy. The most recent time-resolved spectroscopic study in the blue wavelength range was presented
almost 35 years ago by \citet{YSS}. In particular, they found that the variability of the Balmer emission
lines cannot be explained by the canonical hotspot model and that the semi-amplitude $K{_1}$ of their
radial-velocity variations is about 115 \kms. Yet the radial velocity curves are shifted by 30\degr\
relative to the eclipse. \citet{Horne91} concluded in their work that the $K{_1}$ velocity of 115 \kms\
is unreliable and predicted $K{_1}$ to be 58$\pm$11 \kms. This has been neither confirmed nor denied until
now. There was one more spectroscopic study of HT~Cas performed by \citet{MarshHT} with the low-resolution
data. \citeauthor{MarshHT} was able to detect absorption lines from the M5.4$\pm$0.3 secondary star and
to measure its radial velocity semiamplitude $K{_2}$=430$\pm$25 \kms. However, no attempt was made
to study the emission lines in detail and to derive $K_1$.

Among other peculiar characteristics of HT~Cas, there is an unusually small radius of the accretion disc
$R_{d}$ as inferred by many researchers \citep{Zhang, Horne91, Vrielmann2002, Ultracam}. A typical value
of $R_{d}$ in quiescence was measured as being $\sim$0.23$a$. This is only a bit larger than the
circularization radius (0.195$a$, \citealt{VerbuntRappaport}), which determines the theoretically allowed
minimal accretion disc radius. We note that most measurements of $R_{d}$ were based on the
position of the hotspot. In Section~\ref{Sec:HotSpot} we show that
the position of the hotspot does not always give reliable estimates for the accretion disc radius.

\begin{table*}
\caption{Parameters of the most prominent emission lines in the averaged spectra of HT Cas}
\label{Tab:LineParam}
\centering
\begin{tabular}{llcccccccc}
\hline\hline
Set  &Spectral       &  Flux  & Relative &  FWHM  & Peak-to-peak  & \multicolumn{3}{c}{Model parameters}\\
     & line          &(\ergs) & intensity&(\kms) &    (\kms)     & $V_{\rm out}$  (\kms) &  $b$  &  $r_{\rm in}/r_{\rm out}$\\
\hline
1986 & \Halpha       &        & 3.50     & 2130   & 1120          &  590$\pm$5  &  1.59$\pm$0.03   & 0.06$\pm$0.01 \\
     & \Hbeta        &        & 2.75     & 2865   & 1370          &  620$\pm$21 &  2.03$\pm$0.08   & 0.07$\pm$0.01 \\
     & \Hgamma       &        & 2.10     & 3030   & 1450          &             &                  &               \\
     & \Hdelta       &        & 1.89     & 3200   & 1750          &             &                  &               \\
     & \HeI\ \l4471  &        & 1.28     & 3150   & 2090          &             &                  &               \\
     & \HeI\ \l4922  &        & 1.36     &        & 1850          &             &                  &               \\
     & \HeI\ \l5015  &        & 1.28     & 3500   & 2030          &             &                  &               \\
     & \FeII\ \l5169 &        & 1.20     &        &               &             &                  &               \\
     & \HeI\ \l5876  &        & 1.50     & 2860   & 1500          &  690$\pm$9  &  2.36$\pm$0.03   & 0.08$\pm$0.01 \\
     & \HeI\ \l6678  &        & 1.40     & 2150   & 1080          &             &                  &               \\
\hline
1992 & \Hbeta        &        & 2.68     & 2720   & 1360          &  742$\pm$8  &  1.42$\pm$0.08   & 0.06$\pm$0.01 \\
     & \Hgamma       &        & 2.30     & 2800   & 1570          &             &                  &               \\
     & \Hdelta       &        & 2.20     & 2990   & 1520          &             &                  &               \\
     & \HeI\ \l4471  &        & 1.27     &        & 2010          &             &                  &               \\
     & \HeI\ \l4922  &        & 1.34     &        & 1700          &             &                  &               \\
\hline
1995 & \Hbeta        &        & 2.75     & 2440   & 1250          &             &                  &               \\
     & \Hgamma       &        & 2.32     & 2830   & 1670          &             &                  &               \\
     & \Hdelta       &        & 2.08     & 2670   & 1264          &             &                  &               \\
     & \HeI\ \l4471  &        & 1.29     &        &               &             &                  &               \\
     & \HeI\ \l4922  &        & 1.34     &        &               &             &                  &               \\
     & \HeI\ \l5015  &        & 1.36     &        &               &             &                  &               \\
     & \FeII\ \l5169 &        & 1.18     &        &               &             &                  &               \\
\hline
2005-n1& \Halpha     &8.7\tim{-14}&  5.55     & 1960   & 1070          &  572$\pm$6  &  0.81$\pm$0.04   & 0.04$\pm$0.01 \\ 
     & \Hbeta        &7.4\tim{-14}&  3.70     & 2160   & 1090          &  571$\pm$11 &  1.58$\pm$0.08   & 0.09$\pm$0.01 \\
     & \HeI\ \l4922  &6.6\tim{-15}&  1.21     &        & 1520          &             &                  &               \\
     & \HeI\ \l5015  &6.9\tim{-15}&  1.20     & 2340   & 1260          &             &                  &               \\
     & \FeII\ \l5169 &5.0\tim{-15}&  1.19     &        & 1120          &             &                  &               \\
     & \HeI\ \l5876  &2.1\tim{-14}&  1.77     & 2130   & 1160          &  606$\pm$7  &  1.41$\pm$0.04   & 0.02$\pm$0.01 \\
     & \HeII\ \l4686 &4.7\tim{-15}&  1.15     &        & 1640          &             &                  &               \\
\hline
2005-n2& \Halpha     &1.03\tim{-13}& 5.65     & 1960   & 920           &  581$\pm$8  &  0.95$\pm$0.09   & 0.06$\pm$0.02 \\
     & \HeI\ \l6678  &1.0\tim{-14}&  1.42     & 2150   & 1080          &             &                  &               \\
     & \HeI\ \l7064  &8.8\tim{-15}&  1.35     &        & 1140          &             &                  &               \\
\hline
\end{tabular}
\end{table*}

\section{Observations and data reduction}
\label{Sec:Obs}

The spectra presented here were obtained during four observing runs in 1986, 1992, 1995, and 2005.
To check the photometric state of HT~Cas during each set of spectroscopic observations, we also
obtained  several photometric measurements on accompanying telescopes. These observations indicate
that, although HT~Cas has always remained in quiescence, its brightness changed substantially. During
the observations in 1986 and 1992, the V magnitude was about 16.2 mag, whereas in 1995 and 2005 it was
$\sim$17.0 and $\sim$16.8 mag, respectively (Fig.~\ref{Fig:AAVSO}).

The first three observations were conducted in 1986, 1992, and 1995 at the Special Astrophysical
Observatory of the Russian Academy of Sciences, using a 1024-channel television scanner mounted on the
SP-124 spectrograph at the Nasmyth-1 focus of the 6-m telescope \citep{Drabek}. The observations in 1986
were taken during two consecutive nights of September 8 and 9,
about 20 months after the superoutburst. The spectra were taken in the wavelength ranges of 3600--5210
{\AA} (blue spectra) and 5410-7260 {\AA} (red spectra) respectively with a dispersion
of 1.9~{\AA} channel$^{-1}$. The corresponding spectral resolution was about 4.5~{\AA}.
A total of 15 blue and 16 red spectra were obtained with 300 and 360
sec individual exposures. About 1.0 and 1.25 orbital cycles were covered each night.
The observations on December 16, 1992 and  September 9, 1995 were conducted with the same dispersion
in only the blue wavelength range  (4000--5020~\AA\ and 3900--5780~\AA, respectively). 14 and 58
spectra, with 300 and 180 sec individual exposures, were taken during these runs, which covers about 0.8
and 2.0 orbital cycles. We note that the 1995 observations were performed about two months before the normal
outburst, which was particularly well observed by \citet{Ioannou}. Because of electronic focus issues
with the TV scanner during the 1995 observations, the spectral resolution of these data degrades rather
suddenly toward longer wavelengths, especially beyond $\sim$4500 \AA. However, the quality of the spectra
in shorter wavelengths is still reasonably good and suitable for  analysis. Comparison spectra of a HeNeAr
lamp were used for the wavelength calibration. No attempt at flux calibration was made for these data.
The data reduction of the SAO observations was carried out using the procedure described by \citet{Kniazev}.
Hereafter we will refer to these data sets as set-1986, set-1992, and set-1995.

Further observations were obtained in 2005 during two nights of October 29 and 31 at the Observatorio
Astronomico Nacional (OAN SPM) in Mexico on the 2.1-m telescope with the Boller \& Chivens spectrograph,
which was equipped with a 24~$\mu$m ($1024\times1024$) SITe CCD chip. The observations on the first night were
taken in the wavelength range of 4600--6700 {\AA} with a dispersion of 2.05~{\AA} pixel$^{-1}$ while
the rest of the spectra were obtained in the wavelength range of 6150--7225 {\AA} with a dispersion
of 1.05~{\AA} pixel$^{-1}$. The corresponding spectral resolution was about 4.1 and 2.2~\AA, respectively.
A total of 40 spectra with 293 sec individual exposures were taken each
night, covering exactly two orbital periods. Comparison spectra of a CuHeNeAr lamp taken during the night
were used for the wavelength calibration. Both nights of observations were photometric with the
seeing ranged from 1 to 2 arcsec. To apply an accurate flux correction, two standard
spectrophotometric stars at different airmasses were observed every night. They were selected from
Feige110, HR3454, and G191-B2B \citep{Oke:1990aa}. The data reduction was performed using the \IRAF\
environment. We will refer to these data as set-2005, sometimes dividing it into
set-2005-n1 and set-2005-n2 for the first and second nights of observations, if
necessary. A log of observations is presented in Table \ref{ObsTab}.

The orbital phases of the spectra were calculated using the linear ephemeris of \citet{Ultracam}.
The uncertainties of this ephemeris in the orbital phase at the time of our observations are negligible.
For the 2005 data, for which the propagated error is larger, it is less than $2\times10^{-4}$.

\begin{table*}
\caption{Equivalent widths of HT Cas}
\label{Tab:EW}
\centering
\begin{tabular}{lccccccccccccc}
\hline\hline
Set     & Ref. &\Halpha&\Hbeta&\Hgamma&\Hdelta&\l4471&\l4922&\l5015&\l5169 &\l5876&\l6678&\l7064& Magnitude (V) \\
\hline
1980    &  1   &   --   & 77.4 &  51.0 &  41.1 & 11.7 &  9.2 &  9.5 &  --   &  --  &  --  &   -- & 16.0 \\
1981    &  2   &  233.0 & 98.0 &  72.8 &  61.7 & 16.0 & 11.0 & 12.0 & 12.0  & 32.0 & 32.0 &   -- & 16.2 \\
1986    &  3   &  125.9 & 82.0 &  47.9 &  37.9 & 10.2 & 15.0 & 14.3 &  7.2  & 27.4 & 17.4 &   -- & 16.2 \\
1988    &  4   &  186.4 &  --  &   --  &   --  &  --  &  --  &  --  &  --   & 34.1 & 17.4 & 16.1 & 16.2 \\
1992    &  3   &   --   & 76.1 &  54.3 &  48.4 & 12.0 & 12.6 &   -- &  --   &  --  &  --  &   -- & 16.2 \\
1995    &  3   &   --   & 69.9 &  50.7 &  39.3 & 14.5 & 13.3 & 16.5 &  8.0  &  --  &  --  &   -- & 17.0 \\
2005    &  3   &  198.5 & 98.0 &   --  &   --  &  --  &  7.5 &  7.9 &  6.2  & 30.4 & 19.7 & 18.2 & 16.8 \\
\hline
\end{tabular}
\tablebib{
                (1) \citet{YSS}: the spectra were taken on 1980 August 1;
                (2) \citet{Williams83}: the spectra were taken on 1981 December 24$-$27;
                (3) This paper;
                (4) \citet{MarshHT}: the spectra were taken on 1988 July 22$-$24.
}
\end{table*}
\begin{table*}
\caption{The Balmer and \HeI\ decrements}
\label{Tab:Decrement}
\centering
\begin{tabular}{ccccccccc}
\hline\hline
Set     & Ref. &\Halpha/\Hbeta&\Hgamma/\Hbeta&\Hdelta/\Hbeta& \HeI\ \l6678/\l4922&\HeI\ \l5876/\l4471&\HeI\ \l5876/\l6678&\Hbeta/\HeI\ \l4922  \\
\hline
1980    &  1   &     --       &     0.66     &      0.53    &       --           &         --        &          --       &         8.37        \\
1981    &  2   &     2.38     &     0.74     &      0.63    &      2.91          &        2.00       &         1.00      &         8.91        \\
1986    &  3   &     1.54     &     0.58     &      0.46    &      1.16          &        2.69       &         1.57      &         5.47        \\
1988    &  4   &     --       &      --      &       --     &       --           &         --        &         1.96      &          --         \\
1992    &  3   &     --       &     0.71     &      0.64    &       --           &         --        &          --       &         6.03        \\
1995    &  3   &     --       &     0.72     &      0.56    &       --           &         --        &          --       &         5.26        \\
2005    &  3   &     2.03     &      --      &       --     &      2.63          &         --        &         1.54      &        13.07        \\
\hline
\end{tabular}
\tablebib{
                (1) \citet{YSS};
                (2) \citet{Williams83};
                (3) This paper;
                (4) \citet{MarshHT}.
}
\end{table*}

\section{Data analysis and results}
\subsection{Averaged spectra and their long-term variability}
\label{Sec:Decrement}

The averaged and continuum-normalised out-of-eclipse spectra of HT~Cas, uncorrected for orbital motion,
are shown in Fig.~\ref{Fig:aver-spec2}. Here and elsewhere in this paper, out-of-eclipse phases are
defined as 0.1$\leq$$\varphi$$\leq$0.9.
The averaged spectra are similar in appearance to those presented by \citet{YSS} and \citet{Williams83}.
They are dominated by extremely strong and broad double-peaked emission lines of the
Balmer series. Apart from hydrogen, numerous weaker emission lines of neutral helium and singly ionized
iron (\FeII) are present. Also, the high excitation line of \HeII\ \l4686 is clearly detected. Table~\ref{Tab:LineParam} outlines different parameters of the most prominent
lines that were measured from the averaged spectra. In Tables~\ref{Tab:EW} and \ref{Tab:Decrement}, we also
separately present the equivalent width (EW) measurements, including those available in the literature,
and the corresponding Balmer and neutral helium decrement values, respectively.

A comparison of the averaged spectra from different data sets shows significant quantitative differences
between them. There are notable variations in both emission-line strengths and their ratios for different
lines. For example, the EW of \Halpha\ in the set-1986 is $\sim$126~\AA, but in the set-2005 it reaches
almost 200~\AA. The Balmer decrement, being relatively flat in the sets 1992 and 1995, appears rather
steep in the set-1986 and especially in the set-2005, indicating optically thin conditions
in the outer parts of the accretion disc. The decrement
within various series of neutral helium lines (e.g., the singlets \l6678/\l4922 and triplets \l5876/\l4471),
the relative strengths of the \HeI\ triplet and singlet lines (e.g., \l5876/\l6678) and the ratio of
hydrogen to \HeI\ strengths have also changed substantially (Table~\ref{Tab:Decrement}).
This implies that the opacity and the optical thickness of the disc have varied over time, but these
variations do not seem to correlate directly with the system flux (see the last column in
Table~\ref{Tab:EW}).

\begin{figure*}
  \resizebox{\hsize}{!}{
  \includegraphics{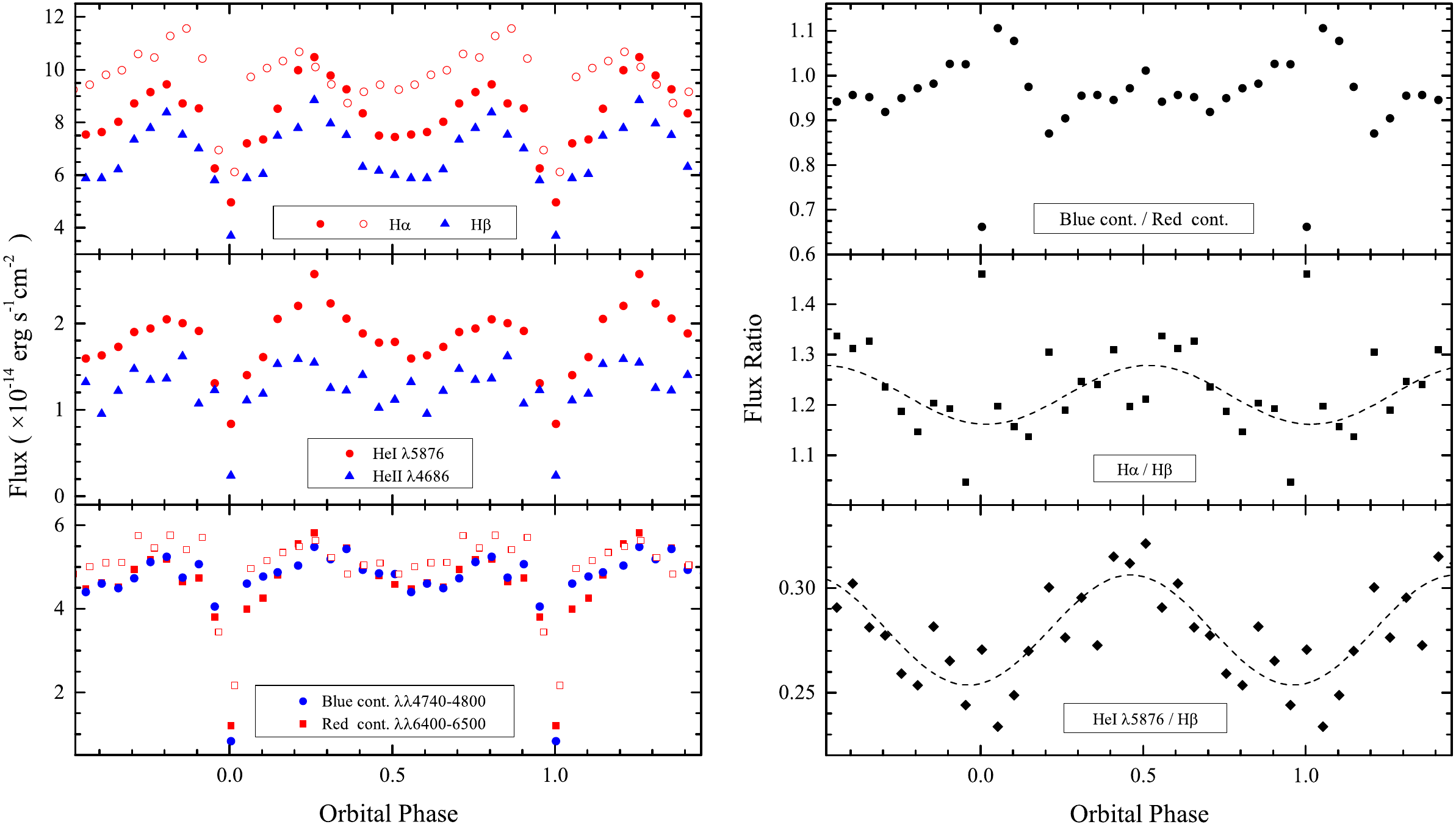}}
  \caption{Left: continuum and emission-line light curves. The filled symbols represent the data from the set-2005-n1,
           the open symbols are for the set-2005-n2. Right: continuum and emission-line flux ratios.}
  \label{Fig:lightcurves}
\end{figure*}

\begin{figure}
\begin{center}
\centerline{
  \includegraphics[width=8.5cm]{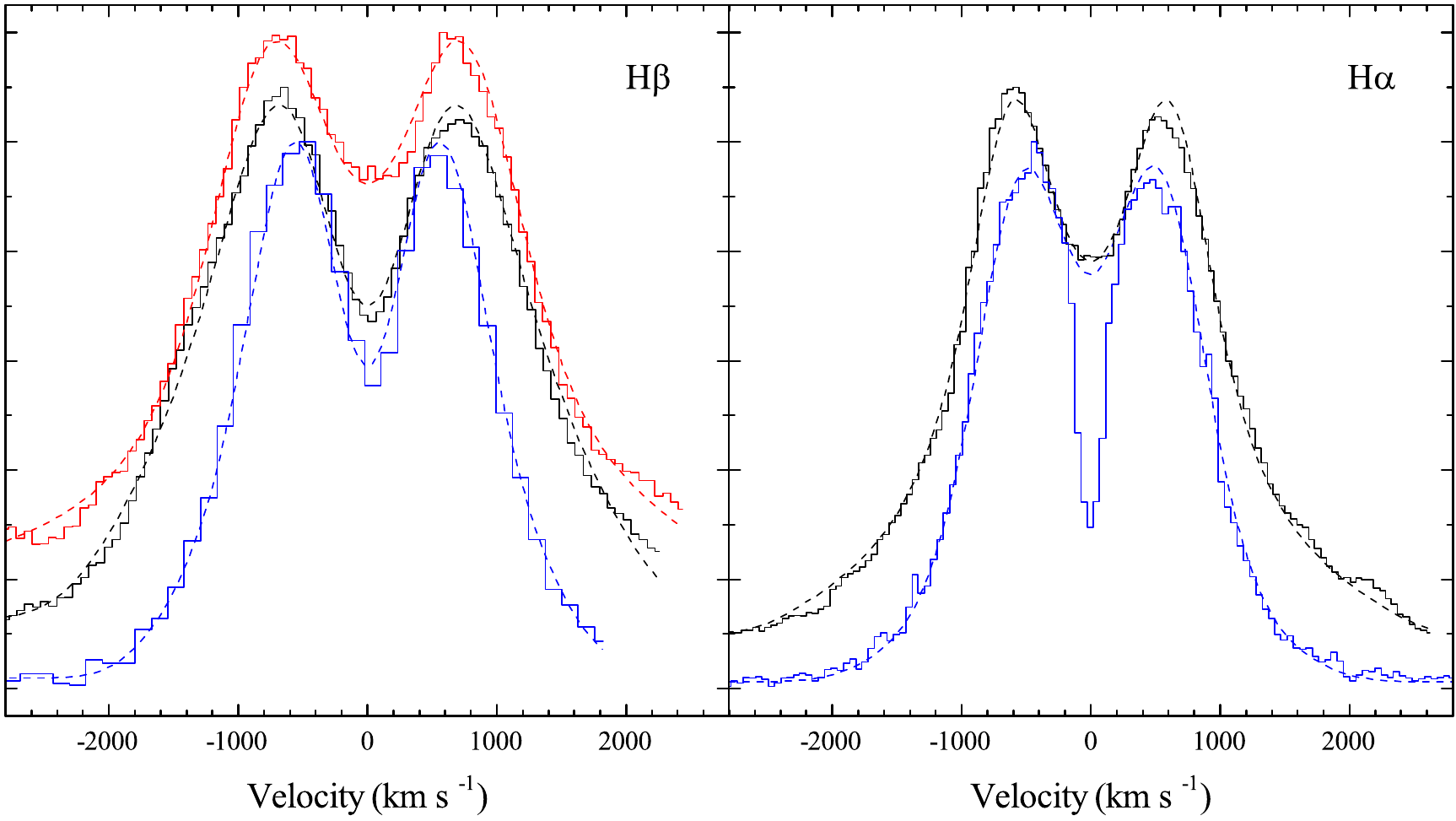}
}
\end{center}
\caption{Averaged profiles of the \Halpha\ and \Hbeta\ emission lines, observed in 1992, 1986, and
2005 (red, black, and blue lines, respectively) together with the corresponding model fits (dashed lines).
The 1992, 1986, and 2005 profiles are shifted vertically by 10\% to prevent overlap.}
\label{Fig:Profiles}

\begin{center}
\centerline{
  \includegraphics[width=8.5cm]{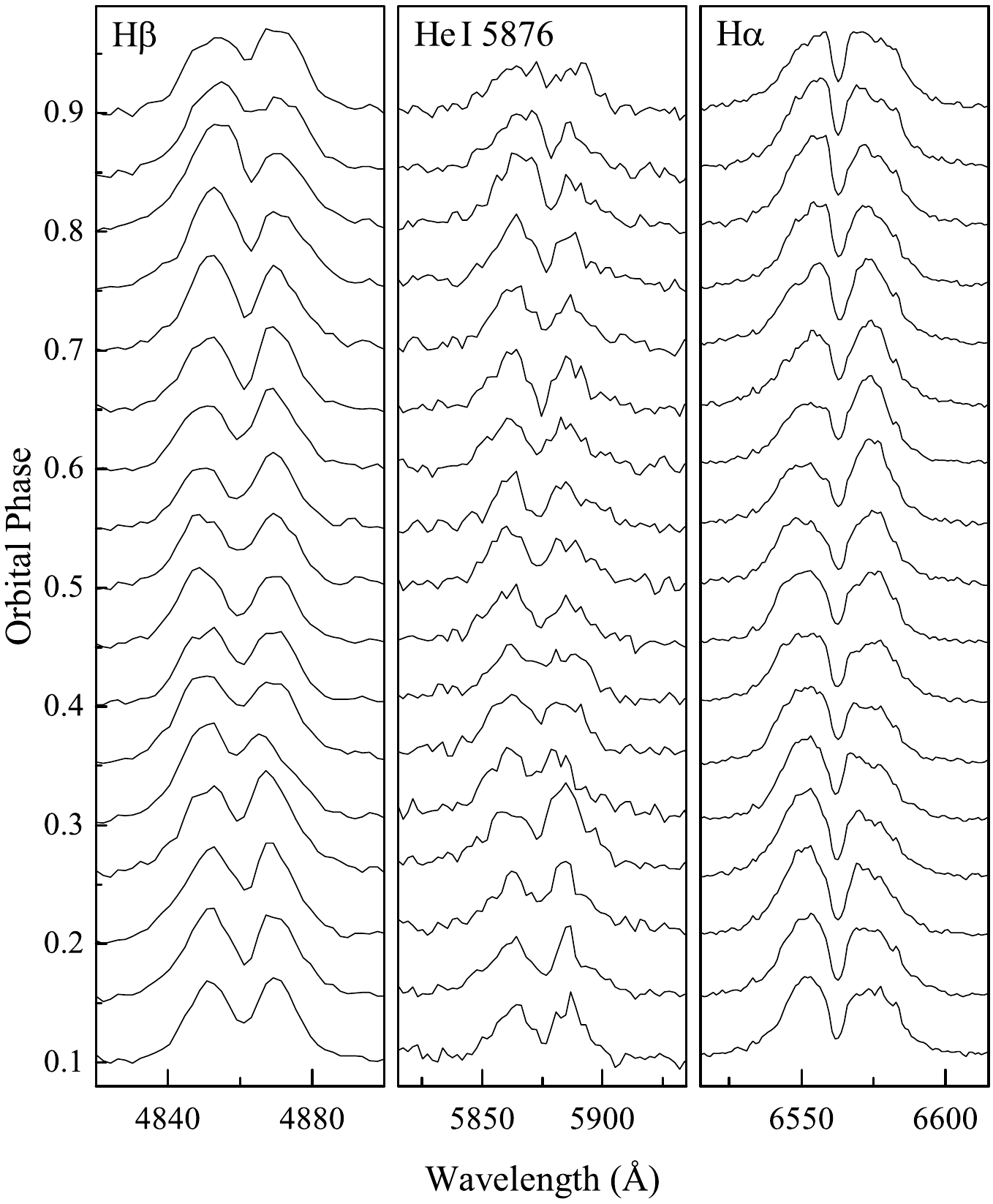}
}
\end{center}
\caption{Evolution of the out-of-eclipse emission-line profiles during the set-2005-n1
         (\Hbeta\ and \HeI\ 5876) and the set-2005-n2 (\Halpha). The spectra have been ordered according
         to phase.}
\label{Fig:OrbProfiles}

\end{figure}

\subsection{Light and colour curves}

The spectra from the set-2005 were used to construct light curves.
The blue and red continuum light curves were computed by summing the flux in the wavelength ranges
\l\l4740--4800 and \l\l6400--6500 \AA, respectively.
The emission-line light curves for \Halpha, \Hbeta, \HeI\ \l5876, and \HeII\ \l4686 were computed by
summing the continuum-subtracted flux inside of $\pm$2700 \kms\ window that was centred at the emission-line
wavelengths. The resulting light curves are plotted in Fig.~\ref{Fig:lightcurves} (left-hand panel).
Although an insufficient phase resolution of the data does not allow us to investigate the eclipse
profiles in detail, we are still able to see their most distinctive features. The continuum shows a deep
almost symmetrical eclipse, with the blue light curve having a deeper eclipse than the red. There is a weak
sign of an orbital hump around phase 0.8--0.9, which is more easily  seen in the blue light, consistent with the
presence of the hotspot. The orbital hump is stronger in the Balmer and \HeI\ lines. The eclipses of
the emission lines have a different shape from the continuum, exhibiting a distinctive shoulder during
egress. The latter feature is seen in both the Balmer, \HeI, and \HeII\ lines, except for the \Halpha\
line from the set-2005-n2. We note that even though the continuum flux was nearly the same during two
nights of the 2005 observations, the \Halpha\ line appeared to be slightly stronger on the second night.

Figure~\ref{Fig:lightcurves} (right-hand panel) also shows the flux ratios of the continuum segments
and Balmer and \HeI\ emission lines. The ratio of the blue and red continuum fluxes (an equivalent of a
colour index) shows little variation outside of eclipses, but there is a sign of an orbital hump with
maximum at phase $\sim$0.95, during which the continuum appears bluer. That is consistent with the
presence of the hotspot. The binary becomes much redder in the middle of the eclipse. The
\Halpha/\Hbeta\ and \HeI\ \l5876/\Hbeta\ ratios show synchronous sinusoidal variations, with the lowest
value also   observed at phase 0.95--1.0. The \Halpha/\Hbeta\ ratio varies by $\sim$10 per cent
around a mean value and there is a notable zigzag jump in the first half of the eclipse. The \HeI\
\l5876/\Hbeta\ ratio varies considerably by a factor of $\sim$2 more than the \Halpha/\Hbeta. However,
the data show no clear evidence for a jump during the eclipse.

\subsection{Accretion disc parameters from modelling of the emission-line profiles}
\label{Sec:modelling}

All emission lines exhibit, in the averaged spectra, slightly asymmetric double-peaked profiles,
with a  blue peak being stronger than the red one. The lines are very broad with a full width at
zero intensity (FWZI) of up to 5--6
thousand \kms\ and a peak-to-peak separation of $\gtrsim$ 1100 \kms\ (in \Halpha), which increases
monotonically towards the higher order Balmer lines. These properties suggest the origin of emission
lines in an accretion disc \citep{Smak1981,HorneMarsh86}. A comparison of the emission lines from
different data sets shows that they vary not only in relative intensity but also in shape and
peak-to-peak separation of the profiles. Figure~\ref{Fig:Profiles} shows the \Halpha\ and \Hbeta\
profiles which exhibit notably different slopes of the line wings.

It is well known that the velocity-separation between peaks in the double-peaked profiles is defined
by the velocity of the outer rim of the accretion disc $V_{out}$ which, in turn, depends on its radius
\citep{Smak1981}. The shape of the line wings is controlled by the surface radial emissivity profile
\citep{Smak1981,HorneMarsh86}, which is commonly assumed to follow a power-law function of the form
$f(r)\propto r^{-b}$, where $r$ is the radial distance from the WD. To estimate accretion disc
parameters, we fitted the averaged emission line profiles using a simple model of a uniform flat
axisymmetric Keplerian geometrically thin disc \citep{Smak1981,HorneMarsh86}.
The primary free parameters of the model are:
\begin{enumerate}
  \item $V_{\rm out}$, the velocity of the outer rim of the accretion disc;
  \item $b$, the power-law index of the line emissivity profile $f(r)$;
  \item $r_{\rm in}/r_{\rm out}$, the ratio of the inner to the outer radii of the disc.
\end{enumerate}

Examples of the application of this technique to the real data are given in
\citet{NeustroevWZ,NeustroevIP,NeustroevSWIFT}. The best-fitting model parameters for the major
emission lines are listed in Table~\ref{Tab:LineParam} and the errors were estimated with a Monte Carlo
approach described in \citet{BorisovNeustroev}. The model fits for the \Halpha\ and \Hbeta\ emission-line profiles from the 1986, 1992, and 2005 data sets are shown in Fig.~\ref{Fig:Profiles} by dashed
lines.

Observations of CVs show that the power-law index $b$ is usually in the range of 1--2, rarely being less
than 1.5 \citep{HorneSaar}. Most of our model fits also give $b$$\approx$1.5--2.0. However, the
best-fitting index $b$ for \Halpha\ from
the set-2005 is $<$1.0. This suggests a flatter radial distribution of the emission-line flux from the
accretion disc of HT~Cas during the 2005 observations. \citet{MarshHorne90} argued that,  to
explain such a behaviour, an increased role of photoionization by the soft X-rays and UV photons from
the centre of the accretion disc should be taken into account. Coupled with the significantly steeper
Balmer and \HeI\ decrements in the 2005 spectra, this may also suggest a lowering of gas density in an
outer area of the disc producing \Halpha\ emission.

\subsection{Emission-line variations and Doppler tomography}
\label{DopMapSec}

\citet{YSS} mentioned that the emission lines of HT~Cas do not vary much in
profile around the orbit (see Fig.~1 in their paper). Our observations confirm this finding as well as
the most mysterious property of profile variations: the blue peak of lines is stronger at phase 0.1--0.2
and the red peak is stronger near phase $\sim$0.6, which is 180\degr\ out of phase than is expected from the ordinary
S-wave (Fig.~\ref{Fig:OrbProfiles}). However, this is correct for the
\Halpha\ and \Hbeta\ lines only, whereas the \HeI\ lines show the opposite, usual behaviour.

\begin{figure*}
   \resizebox{\hsize}{!}{
   \centerline{\includegraphics[width=17.3cm]{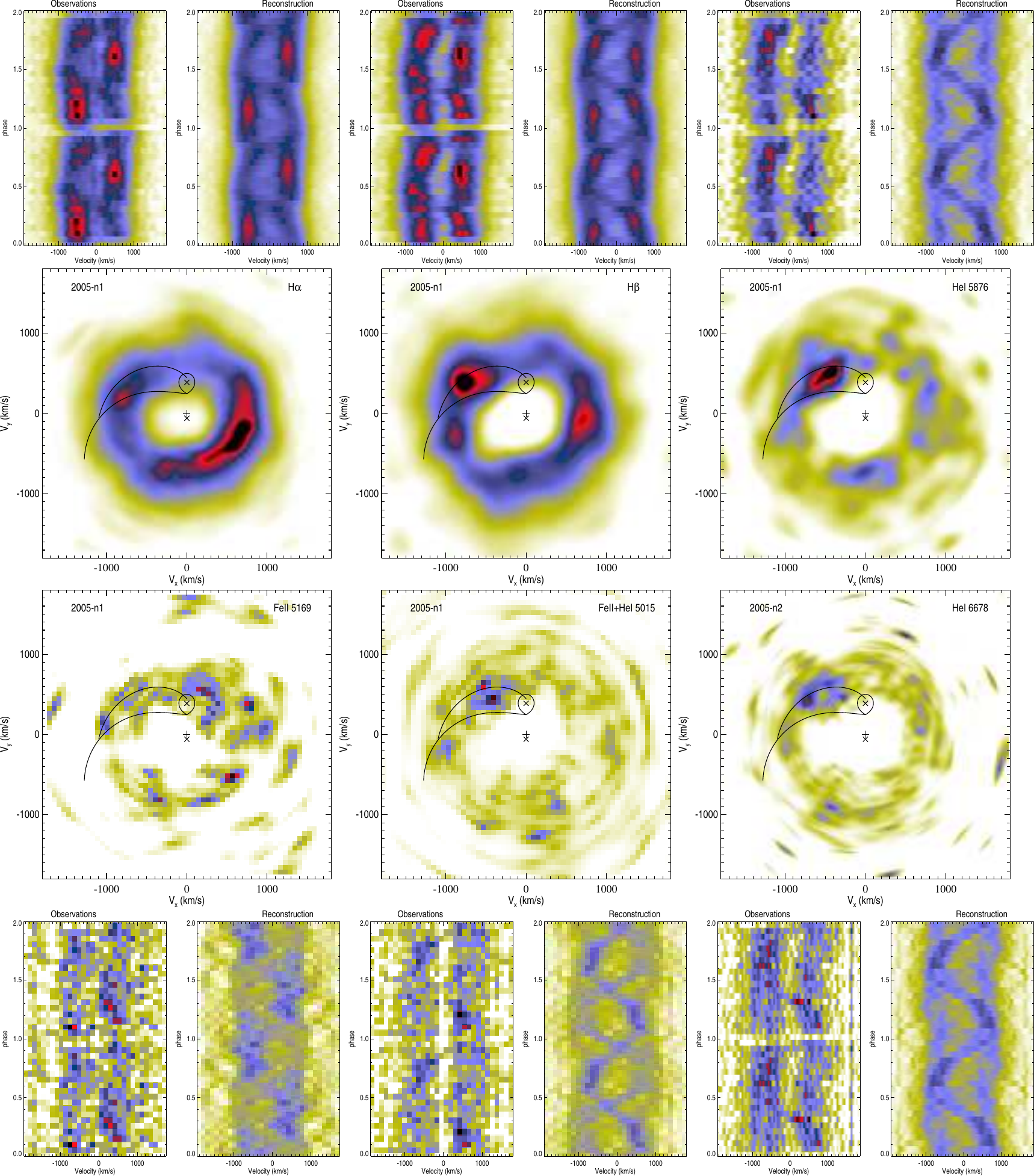}}}
    \caption{Doppler tomography for the \Halpha, \Hbeta,\ and \HeI\ 5876 emission lines (in the upper
             half of the figure), and for \FeII\ 5169, \HeI\ 5015, and \HeI\ 6678 (in the bottom half of
             the figure) from the data sets 2005-n1 and 2005-n2. For each line, the Doppler maps (two
             middle panels) and corresponding observed and reconstructed trailed spectra (top and
             bottom panels) are shown. Indicated on the maps are the positions of the WD (lower cross),
             the centre of mass of the binary (middle cross) and the Roche lobe of the
    secondary star (upper bubble with the cross). The predicted trajectory of the gas stream and the
    Keplerian velocity of the disc along the gas stream have also been shown in the form of the lower
    and upper curves, respectively. The Roche lobe of the secondary and the trajectories have been
    plotted using the system parameters, derived by \citet{Horne91}.}
    \label{Fig:dopmaps2005}
\end{figure*}

More details are revealed in the trailed spectra. In top and bottom panels of Fig.~\ref{Fig:dopmaps2005},
the most representative lines from the data sets 2005-n1 and 2005-n2 are shown. One can clearly recognize
that the ordinary S-wave is certainly present in most of the lines, yet the anomalous emission source
mentioned above is strong in \Halpha, weaker in \Hbeta, and very weak or undetectable in \HeI\ lines.
We note that even though the \HeII\ \l4686 line is clearly seen in averaged spectra, it is still too
weak and noisy to reveal any variability.

\begin{figure*}
    \centerline{\includegraphics[width=17.9cm]{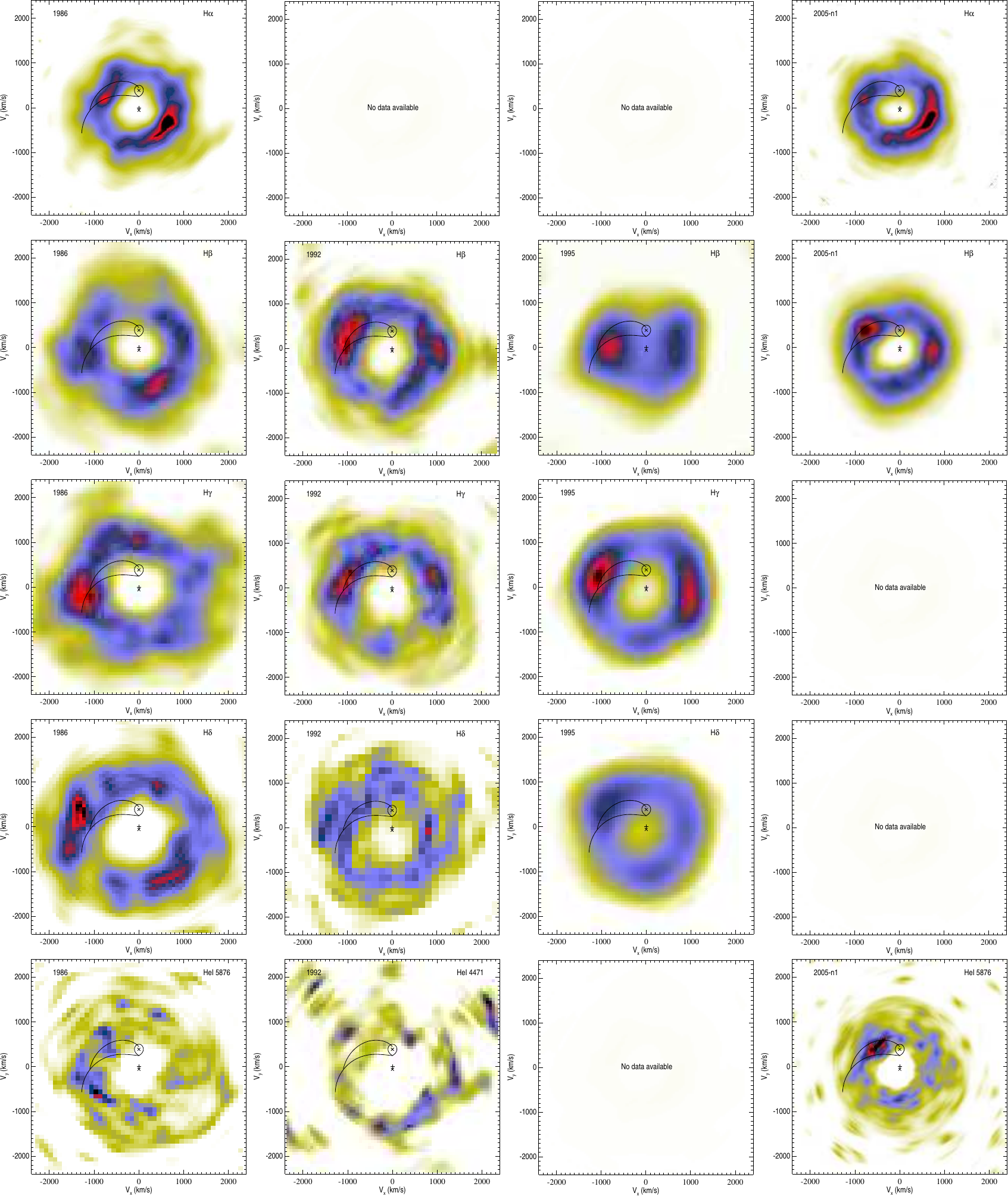}}
    \caption{Doppler tomography for the Balmer and \HeI\ emission lines. Each column shows
             the maps for different data sets (1986, 1992, 1995 and 2005-n1). Each row shows the maps for
             different lines (\Halpha, \Hbeta, \Hgamma, \Hdelta\ and \HeI).}
    \label{Fig:dopmapsAll}
\end{figure*}

To provide a more convincing picture of the sources of emission in the accretion disc of HT~Cas,
we used Doppler tomography \citep{MarshHorne88}. For a comprehensive review of the method and many
examples of its application, see \citet{Marsh2001} and references therein. Figures~\ref{Fig:dopmaps2005}
and ~\ref{Fig:dopmapsAll} show the Doppler maps computed using the code developed by \citet{Spruit}.

\begin{figure}[t]
    \includegraphics[width=3.0cm]{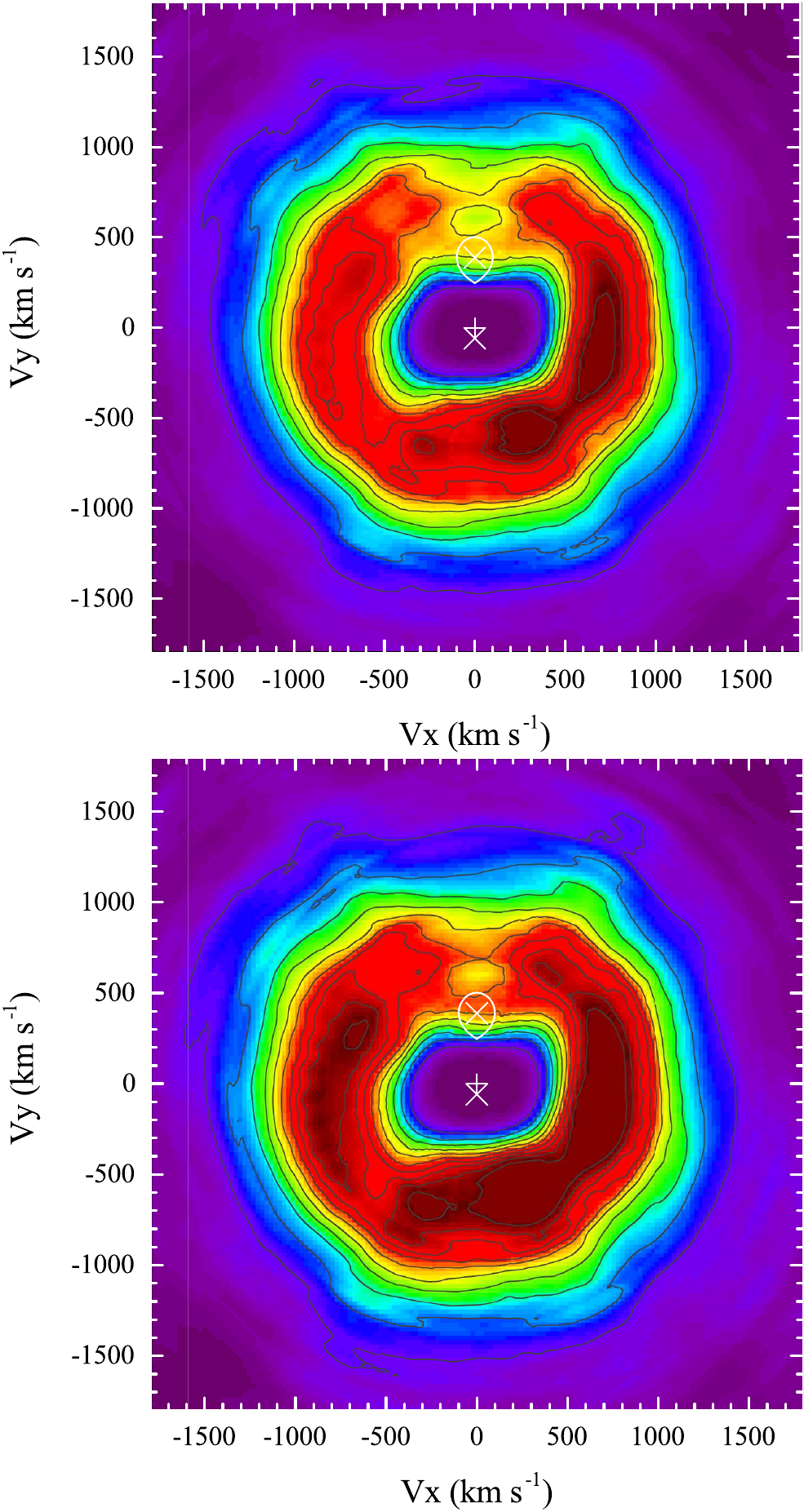}
    \includegraphics[width=5.5cm]{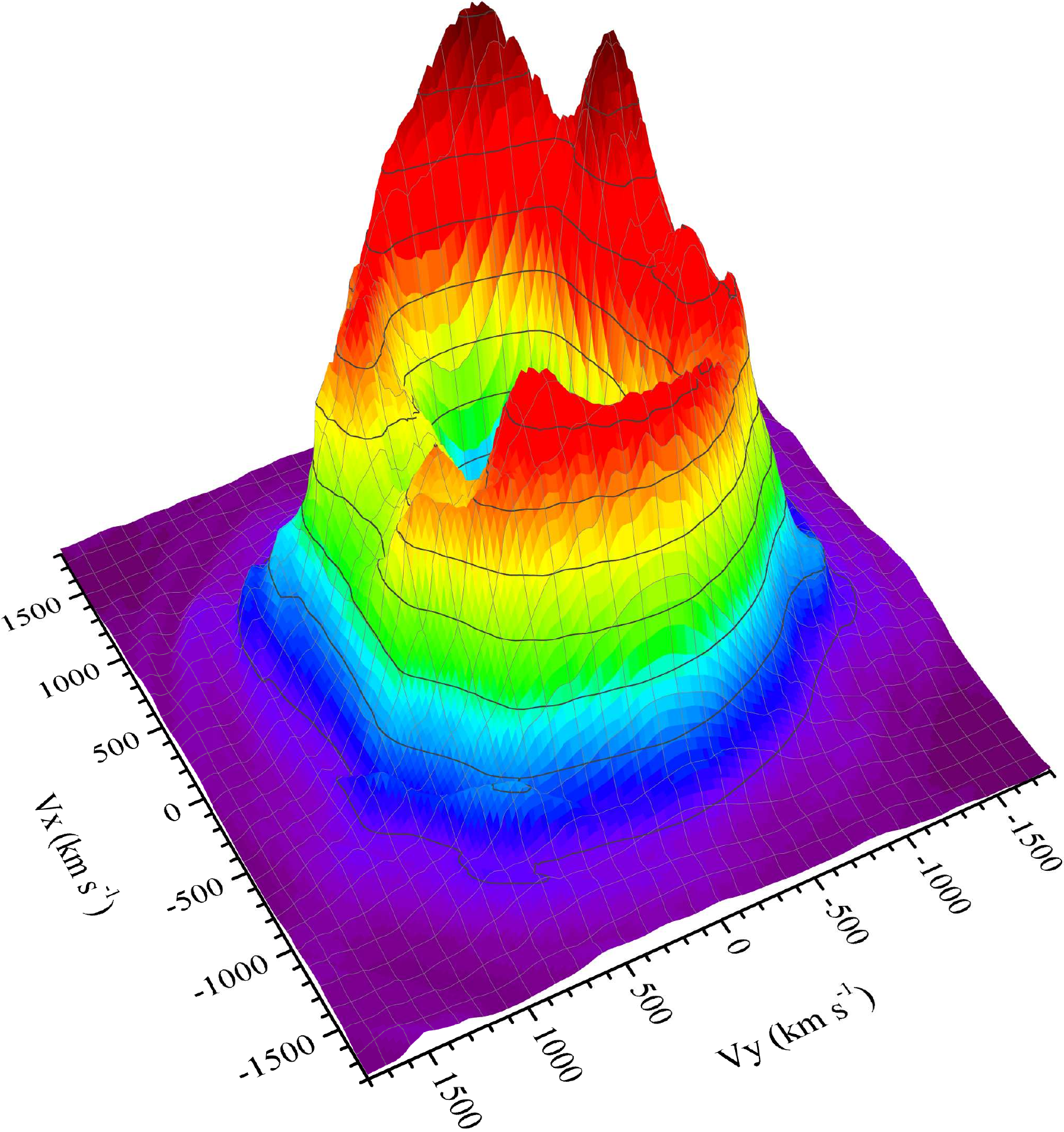}
    \caption{Doppler map of the \Halpha\ emission line from the set-2005-n2 in 3D (right)
             and 2D representations with different contrasts to emphasise various components
             of the map (left).}
    \label{Fig:dopmaps3D}
\end{figure}

We start the discussion with the 2005 data set as these spectra have higher signal-to-noise ratio (S/N) and
spectral resolution and they produce maps of the best quality. The tomograms of the representative lines
from this set are shown in Fig.~\ref{Fig:dopmaps2005}, together with the corresponding reconstructed
counterparts, shown in the top and bottom panels, alongside  the trailed spectra.
Since the gradual occultation of
the emitting regions during eclipse is not taken into account, we constrained our data sets by removing
eclipse spectra covering the phase ranges $\phi = 0.9{-}1.1$.
To help in interpreting the Doppler maps, the positions of the WD, the centre of mass
of the binary  and the Roche lobe of the secondary star
are marked. The predicted trajectory of the gas stream and the Keplerian velocity of the disc along
the gas stream have also been shown in the form of the lower and upper curves, respectively. The Roche
lobe of the secondary and the trajectories were plotted using the system parameters that were derived by
\citet{Horne91}: an inclination $i$=81\degr, mass ratio $q$=0.15, and $M_1$=0.61 \Msun.
We note that the classical two-dimensional visualisation of tomograms used in Fig.~\ref{Fig:dopmaps2005}
does not always provide adequate representation of the multicomponent structure that we discuss below.
We find it useful to also present the Doppler map of \Halpha\ from the set-2005-n2  in three-dimensional
(3D) form (Fig.~\ref{Fig:dopmaps3D}).

All the Doppler maps display a ring of disc emission, the  radius of which is different for various lines,
reflecting the varied peak-to-peak velocity separation in these lines. However, the detailed
appearance of the Balmer and helium tomograms is rather dissimilar. Most of the maps show a compact
emission area in the fourth quadrant ($-V_{x}$,$+V_{y}$), which can be unequivocally identified as
the hotspot that is located in the region of interaction between the gas stream and the outer edge of the
accretion disc. This area is a dominant emission source in \Hbeta\ and \HeI\ lines, but very weak in
\Halpha. Instead, the \Halpha\ map exhibits a bright enhanced emission region in the second quadrant
($+V_{x}$,$-V_{y}$) whose ambiguous nature we discuss in Section~\ref{Sec:LeadingSpot}.
The 3D Doppler maps of highest quality (\Halpha, \Hbeta,
\HeI\ \l5876) clearly show that the hotspot is located on the top of azimuthally extended emission
structure of spiral shape (Fig.~\ref{Fig:dopmaps3D}). This spiral structure in the fourth quadrant
and the emission region in the second quadrant are disjointed from each other by gaps in the upper- and
lower-left parts of the tomograms. The \FeII\ \l5169 line shows no clear evidence for either
of the three emission sources. No sign of the secondary star is seen in either tomogram.

To examine the stability of the emission structure of HT~Cas over time, we calculated Doppler
maps for the four Balmer and one \HeI\ lines from other available data sets (Fig.~\ref{Fig:dopmapsAll}).
Even though the quality of these data is notably worse than of the 2005 data set, they show the same
features in Doppler maps. The main dissimilarity between observations is a different contribution
of the above-mentioned components.

\subsection{Radial velocity study}
\label{Sec:RadVel}

To the best of our knowledge, there was only one attempt to estimate the radial velocity semi-amplitude
of the white dwarf in HT~Cas in the past. \citet{YSS} measured the $K_1$ velocity of $115\pm6$ \kms\ for
the emission lines, but the resulting radial velocity curve is 30\degr\ out of phase with the WD.
\citet{Horne91} showed that this result is inconsistent with most of the photometric data and predicted
$K_1$ velocity of $58\pm11$ \kms.

We measured the radial velocities of the emission lines in HT~Cas by applying the double-Gaussian method
described by \citet{sch:young} and refined by \citet{Shafter1983}. This technique consists of convolving
each spectrum with a pair of Gaussians of width $\sigma$ , the centres of which have a separation of $\Delta$. The
position at which the intensities through the two Gaussians become equal is a measure of the wavelength
of the emission line. The measured velocities depend on the choice of $\Delta$ and, by varying its value,
different parts of the lines can be sampled. It is commonly believed that the most reliable parts of the
emission-line profile for deriving the radial velocity curve are the extreme wings. \citet{YSS} used
$\Delta$=3200 \kms\ in their measurements.

To test for consistency in the derived velocities and the zero phase, we separately used the
lines \Halpha\ and \Hbeta\ in the set-2005-n1, and \Halpha\ in the set-2005-n2. The measurements were
made using the Gaussian $\sigma$ of 100 and 200 \kms\ and different
values of the Gaussian separation $\Delta$ ranging from 1200 \kms\ to 4000 \kms\ in steps of 50 \kms,
following the technique of ``diagnostic diagrams'' \citep{Shafter1986}. For each value of $\Delta$ we
made a non-linear least-squares fit of the derived velocities to sinusoids of the form:
    \begin{equation}  \label{radvelfit}
        V(\varphi,\Delta )=\gamma (\Delta )-K_1(\Delta )\sin \left[ 2\pi \left(
        \varphi-\varphi_0\left( \Delta \right) \right)\right]\,,
    \end{equation}
where $\gamma$ is the systemic velocity, $K_1$ is the semi-amplitude, $\varphi_0$ is the phase of
inferior conjunction of the secondary star, and $\varphi$ is the phase calculated according to the
ephemeris from \citet{Ultracam}. During this fitting procedure we omitted spectra that covers the phase
ranges $\varphi = \pm0.1$, owing to measurement uncertainties during the eclipse.

\begin{figure}[t]
\centering
\includegraphics[width=8.0cm]{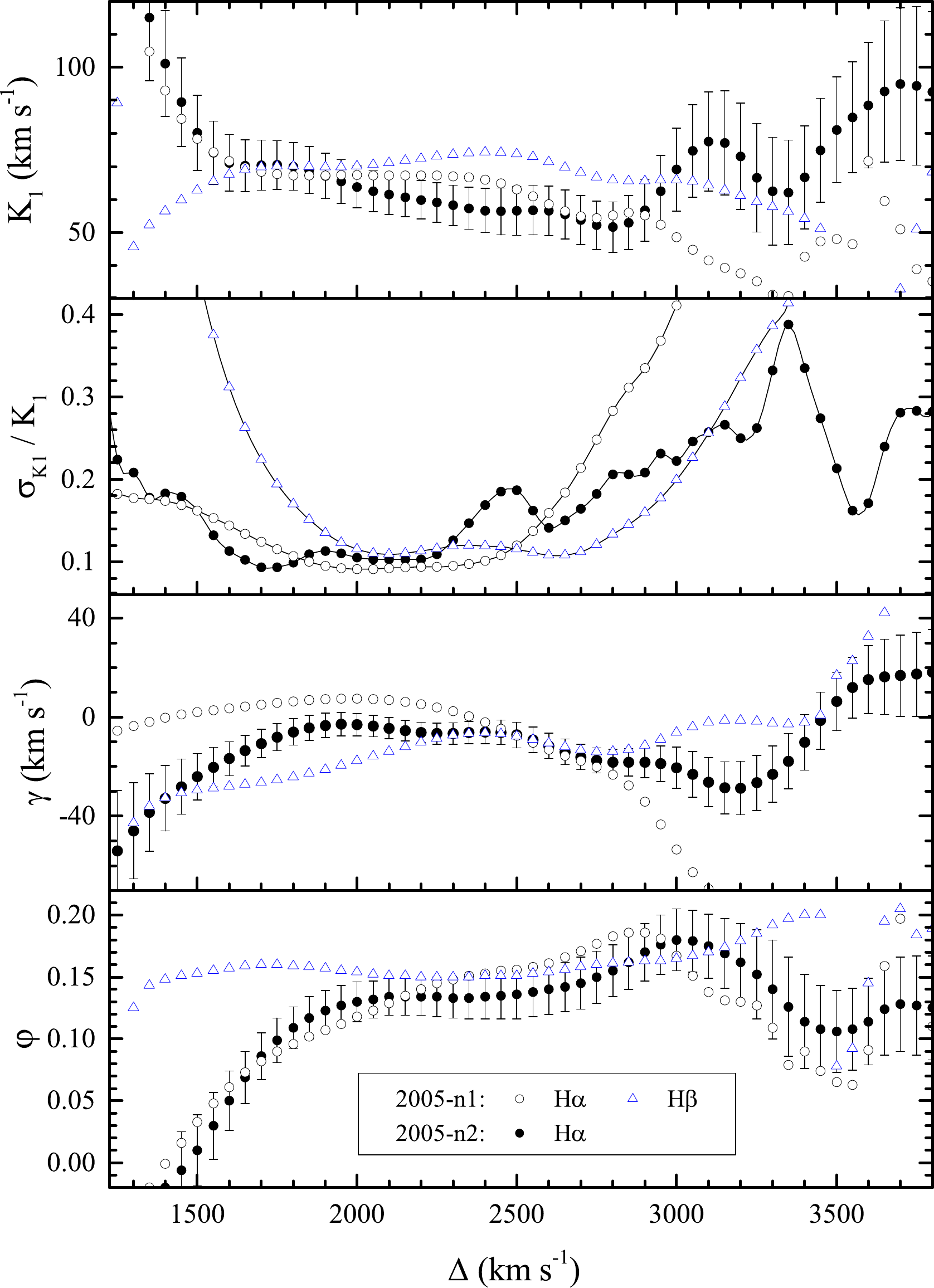}
\caption{The diagnostic diagram for the \Halpha\ and \Hbeta\ emission lines from the sets 2005-n1
         and 2005-n2, showing the response of the fitted orbital elements to the choice of the
         double-Gaussian separation.}
\label{Fig:diagram}
\end{figure}

The resulting diagnostic diagrams are shown in Fig.~\ref{Fig:diagram}. The diagrams show the
variations of $K_1$, $\sigma(K_1)/K_1$ (the fractional error in $K_1$), $\gamma,$ and $\varphi_0$
with $\Delta$ \citep{Shafter1986}. To derive the orbital elements of the line wings, \citet{ShafterSzkody}
suggest  taking the values that correspond to the largest separation $\Delta_{max}$, just before
$\sigma(K_1)/K_1$ shows a sharp increase. We note, however, that all the parameters are consistent
for different lines and are quite stable over a reasonable range of Gaussian separations around
$\Delta_{max}$ which can be set at $\sim$2500--2700 \kms. Using $\sigma(K_1)$, $\sigma(\gamma)$
and $\sigma({\varphi}_0)$ as a weight factor, we find the following mean values
of the orbital patameters: $K_1=61\pm8$ \kms, $\gamma=-9\pm5$ \kms, and ${\varphi}_0=0.15\pm0.02$.

The derived value of $K_1$ is very much consistent with the one predicted by \citet{Horne91}.
However, the radial velocity curves of all the investigated emission lines are significantly shifted
relative to the eclipse ($\sim$55\degr), and therefore these lines cannot be used to represent the motion
of the WD. As seen in the diagnostic diagrams (Fig.~\ref{Fig:diagram}), the shift is observed over the
whole range of separations $\Delta$, from emission-line profile peaks to the extreme wings where the
noise begins to dominate. This shift is almost twice as large as that reported by \citet{YSS}.
Although the Doppler maps of HT~Cas have a very complex structure, they show no evidence for
compact emission or absorption sources at the far wings of spectral lines, which can be responsible for
the observed phase shift in this velocity range ($>$1500 \kms). This fact suggests that a global
asymmetric configuration may exist in the inner parts of the accretion disc of HT~Cas (e.g. eccentric,
elliptical structure).

\section{Discussion}

A Doppler tomography technique applied to a large set of multi-epoch spectroscopic observations of HT~Cas
reveals an unusual distribution of emission in this system. The tomograms
show at least three areas of enhanced emission: the hotspot superposed on the spiral structure in the
fourth quadrant, and the broad, extended bright region of uncertain origin in the second quadrant.

To explain this structure, we need to assume its spatial location in the binary system,
but this is not possible without a detailed knowledge of the velocity field.
Although the standard Shakura-Sunyaev model assumes that the gas in the accretion disc
moves with circular Keplerian velocities \citep{ShakuraSunyaev}, this assumption is somewhat
aprioristic and it is still unclear if the velocity field of accretion discs is actually Keplerian.
Furthermore, this assumption may not be accurate in a strict sense, because some of the system components,
such as the gas stream do not follow a pure Keplerian rotation law \citep{MarshUGem}. The formation of
spiral shocks in the hot discs of outbursting CVs may also cause deviations from Keplerian motion. Indeed,
\cite{Baptista} presented evidence that the velocity of the emitting gas along the spiral pattern in
the accretion disc of IP~Peg in outburst is lower than the Keplerian velocity for the same disc radius.

\begin{figure}
\begin{center}
\centerline{
  \includegraphics[width=8.5cm]{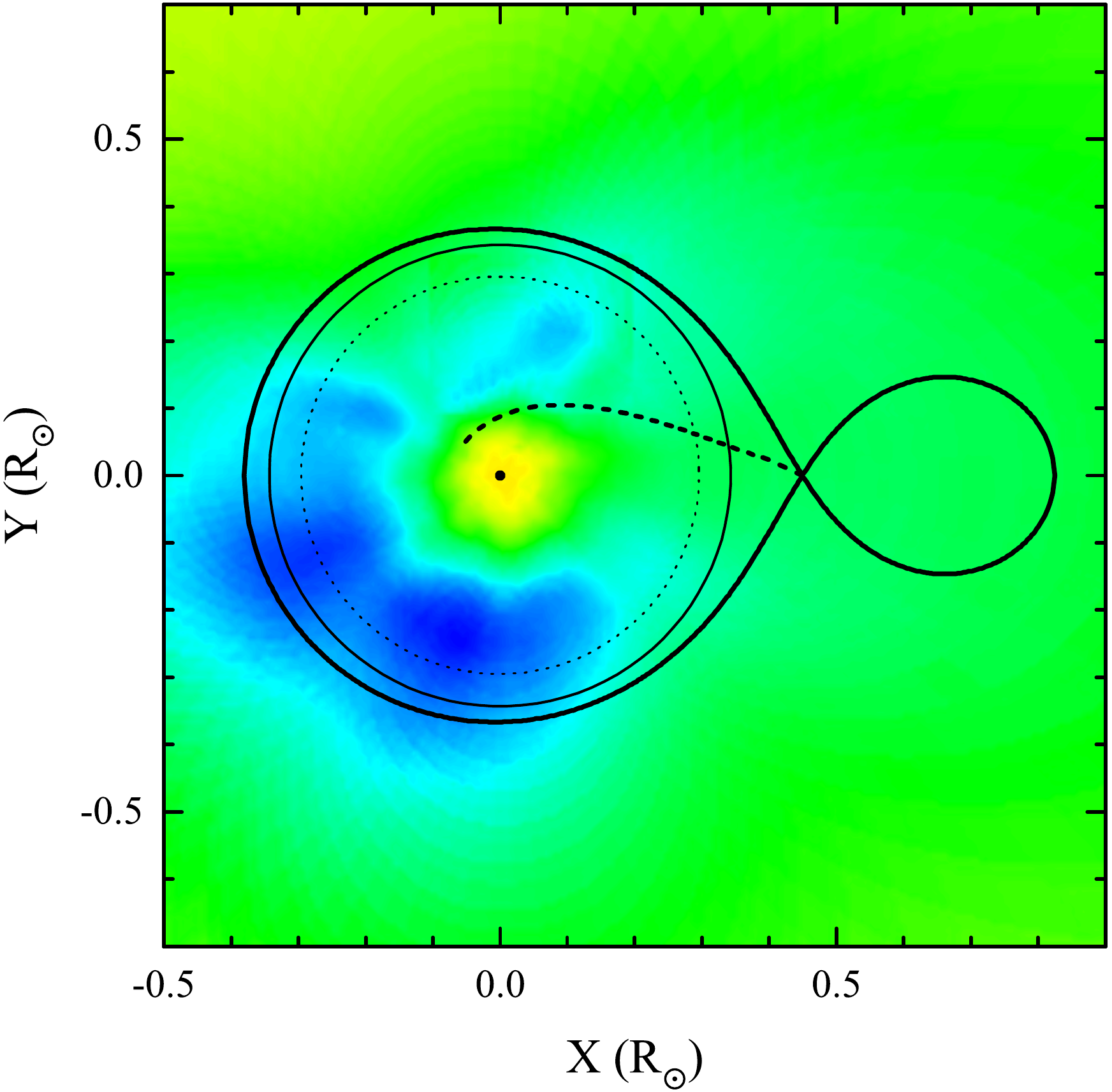}
}
\end{center}
\caption{The \Halpha\ Doppler map from the 2005 observations transformed to a spatial image, assuming
         a circular Keplerian flow in the disc. The Roche lobes of the stars (thick solid line), the
         tidal truncation (thin solid line), and 3:1 resonance (thin dashed line) radii are also shown.
         The thick dashed line represents the gas stream from the donor star.}
\label{Fig:GeoMaps}
\end{figure}

On the other hand, we find that the Keplerian assumption gives an acceptable working basis, since no
significant deviation from the Keplerian rotation has been ever reported for quiescent
accretion discs. Instead, evidence of material in Keplerian rotation has been obtained by many authors.
For example, \citet{Marsh-IPPeg} found strong evidence for Keplerian rotation of the quiescent disc of
 above-mentioned IP~Peg, the only detected deviations being caused by the stream. \citet{Ishioka} also
conclude that the behaviour of the disc of IP~Peg is basically consistent with Keplerian rotation,
though more complex than those predicted by a simple axisymmetric model. In the particular case of
HT~Cas, \citet{YSS} presented a crude verification that the disc in this system is Keplerian over at
least a factor of 4 in radius. Bearing these considerations in mind, we thus  find it useful to create
and show, for illustration purposes and to facilitate further discussion, a map of the \Halpha\ emission
in spatial coordinates, assuming a circular Keplerian flow in the disc (Fig.~\ref{Fig:GeoMaps}).

The accretion disc, as it appeared in Fig.~\ref{Fig:GeoMaps}, looks rather unusual. A general impression
is that the disc of HT~Cas is indeed patchy, as was suggested by \citet{Vrielmann2002}. We discuss the
derived structure in the following subsections.

\subsection{Hotspot and the size of the accretion disc}
\label{Sec:HotSpot}

Assuming the Keplerian velocity in the accretion disc, the measured projected outer disc velocity
$V_{\rm out}$ can be used to determine the radius of the disc:
\begin{equation}
\label{VelKep}
  R_{d} = \frac{G M_1 \sin^{2}i}{V^{2}_{\rm out}}\,.\end{equation}
Individual spectral lines are sensitive to the local physical conditions in the disc (e.g. gas density
and temperature), so different lines may lead to different apparent disc radii, which reflects the region
where that particular line is excited. We measured $V_{\rm out}$ through the modelling of the \Halpha\
emission line that has a relatively low excitation energy and that has been proven to be a good tracer
of the cool outer parts of the accretion disc. If \Halpha\ is not present in the given data set,
we used the \Hbeta\ line (see Section~\ref{Sec:modelling} and Table~\ref{Tab:LineParam}).
Firstly, we used the 2005 data, for which
we adopted the value for $V_{\rm out}$ as being 575$\pm$4 \kms. This value is the weighted mean for the
\Halpha\ and \Hbeta\ lines from the two nights, using $\sigma$($V_{\rm out}$) as a weight factor.
Using the system parameters derived by \citet{Horne91}, this value gives $R_{d}$$\approx$0.52$\pm$0.01$a$.

The obtained disc radius coincides precisely with the tidal truncation radius of the accretion disc
$r_{\rm max}$=0.522$a$ (for $q$=0.15), which can be estimated using Equation 2.61 from \citealt{Warner:1995aa})

\begin{equation}
\label{RmaxA}
   r_{\rm max}  =a  \frac{0.6}{1+q}\,.
\end{equation}

\noindent

Such a large disc is slightly tidally distorted and elongated perpendicular to the line of
centres of the WD and the secondary star. This means that some deviations from circular Keplerian
flow are expected at the outer disc. The magnitude of these deviations were estimated by different
methods with roughly consistent results. For example, \citet{SteeghsStehle} showed, using 2D
hydrodynamic calculations, that the departures can reach 100 \kms, which is in agreement with the
inviscid (non-interacting single particle) calculations of \citet{Paczynski}. Because of this effect,
the peak-to-peak separation of emission-line profiles is also expected to vary as a function of binary
phase, which affects the disc radius measurement. This variation can, in principle, be measured using
time-resolved spectroscopy. Unfortunately, we  failed to either confirm or deny this effect with
our medium-resolution data. Individual line profiles in our spectra are relatively noisy and they are
significantly affected by small- and large-scale structures in the accretion disc, which are difficult
to take into account. However \citet{SteeghsStehle} pointed out  that, for
the orbit averaged spectra, the velocity deviations mostly cancel out, so the assumption of circular Keplerian
velocities is still reliable. Nevertheless, to be on the conservative side, we assume that the
average disc size measured from the orbit averaged profiles is accurate to better than 10\%. Thus, for
HT~Cas we adopt $R_{d}$=0.52$\pm$0.05$a$.

\begin{figure*}
\centerline{
  \includegraphics[height=8cm]{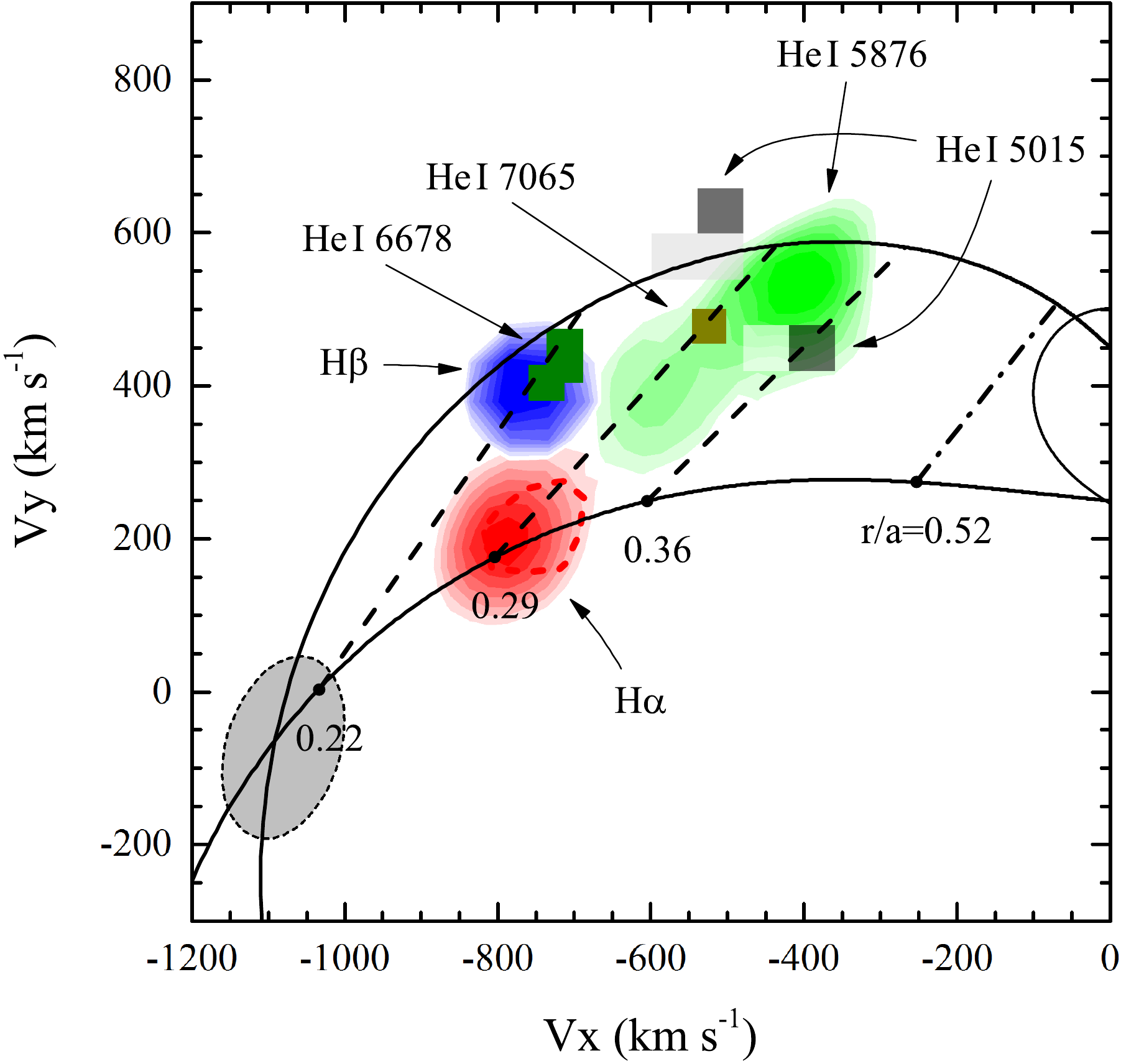}
  \includegraphics[height=8cm]{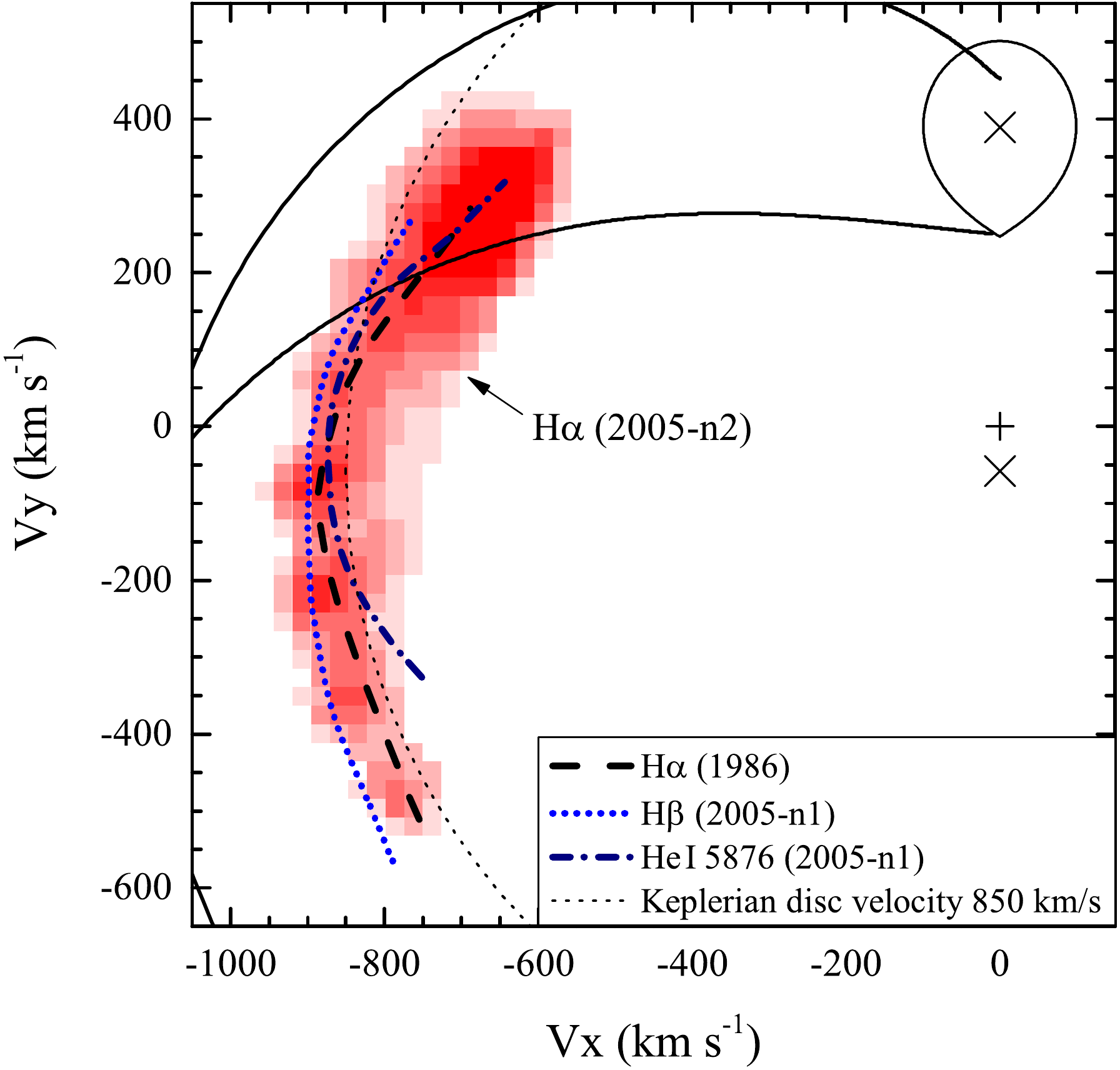}
}
\caption{Left: Zoomed hotspot area of the Doppler map combined from all the tomograms taken during the 2005
         observations, with different contrasts to emphasise the hotspot location. The map also shows
         the spot location from the 1986 observations. The \Halpha\ spot is marked by the red dashed
         line. A grey oval shows the other spectral lines. The dashed lines connect the velocity of
         the ballistic gas stream (lower curve) and the velocity on the Keplerian disc along the gas
         stream (upper curve) for the same points at distances labelled along the lower curve (in $r/a$
         units). The dash-dotted line corresponds to the measured radius of the accretion disc, which
         coincides with the tidal truncation radius $r_{\rm max}/a$=0.52. Right: Zoomed part of the
         \Halpha\ Doppler map from the set-2005-n2 centred around the spiral arm area. The thick lines
         show the trace of this area in other spectral lines and data sets. Dotted thin line shows the
         Keplerian disc velocity of 850 \kms.
         }
\label{Fig:BrightSpot1}
\end{figure*}

A comparison of the peak-to-peak separation of emission lines from different observations, including
those presented by \citet{YSS}, shows that such a large disc seems to be normal for HT~Cas. Most of
the data (except for the 1992 observations) suggest that the accretion disc radius $R_{d}$ lies within
the range 0.45$-$0.52$a$. This result is not consistent with previous radius measurements. As we
mentioned in Sect.~\ref{Sec:HT_Cas}, a typical value of $R_{d}$ in quiescence was measured as
just $\sim$0.23$a$. We note that most of the measurements  found in the literature
were based on the position of the hotspot. Among the methods used were the eclipse-mapping technique
\citep{Ultracam} and methods based on contact timings of the bright spot eclipses \citep{Horne91}, and
on modelling the eclipse data \citep{Zhang}.
Our data, however, indicate that the hotspot is not located at the accretion disc edge.

Figure~\ref{Fig:BrightSpot1} (left-hand panel) shows a zoomed hotspot area of the map, which combines  all
the tomograms from the 1986 and 2005 observations. The position of the hot spot in all the emission lines
from 2005 is consistent with the trajectory of the gas stream and either has a velocity close to  the
expected velocity of the stream at a distance from the WD (i.e., \Halpha), or a Kepler velocity at the
stream position (\Hbeta, \HeI\ 6678), or a mix of these velocities (\HeI\ 5876). In 1986 the \Halpha\
spot is seen in nearly the same position as in 2005, whereas all other lines show the spot (marked by a
grey oval) somewhat further along the stream trajectory. The latter spot is also evident in Doppler maps
from other data sets. It is interesting that all the peaks of spot emission are located much closer to
the WD ($R_{hs}$$\approx$0.22--0.30$a$) than the disc edge that was measured from the double-peaked profiles
($R_{d}$$\approx$0.52$a$). This suggests that the gas stream flows almost unaffectedly through the outer
disc regions before it starts to be seen as a continuum and line-emission source.
A similar finding is reported by \citet{WZ1} and \citet{WZ2} for the dwarf nova \object{WZ~Sge}.

Such a low density outer disc region is expected to have much lower optical depth than the inner disc,
which is in agreement with the rather steep Balmer decrement to be observed in HT~Cas. We also note the work
of \citet{Vrielmann2002} who perform a Physical Parameter Eclipse Mapping analysis of multicolour
photometric observations of HT~Cas and show that the accretion disc is moderately optically
thin, but becomes nearly optically thick near the WD. They found that in the $R$ band, the disc region
outside $\sim$0.28$a$, which they call the disc edge, contributes less than 1\% of the flux compared to the WD.
We point out that, in fact, this edge is the photometric one, after which the accretion disc gas does
not produce much broadband continuum light. The spectroscopic data suggest, however, that optically thin
gas, visible in spectral lines, is extended further than the photometric disc. There is also evidence
that such disagreement between photometric and spectroscopic sizes of the accretion disc can be
observed not only in quiescence, but also in outbursts \citep{Isogai2015}.

\begin{figure*}
\centerline{
  \includegraphics[height=8cm]{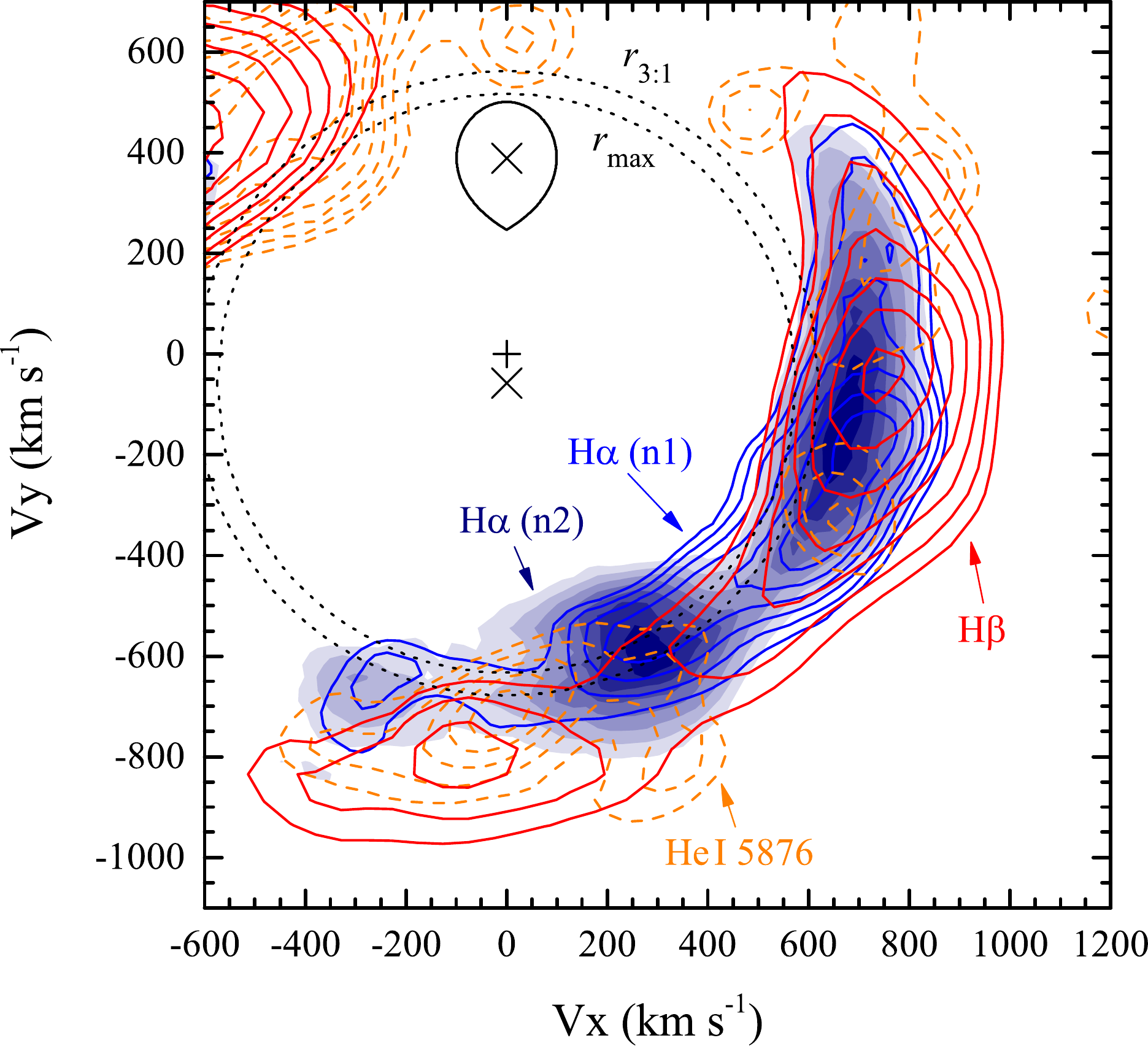}
  \includegraphics[height=8cm]{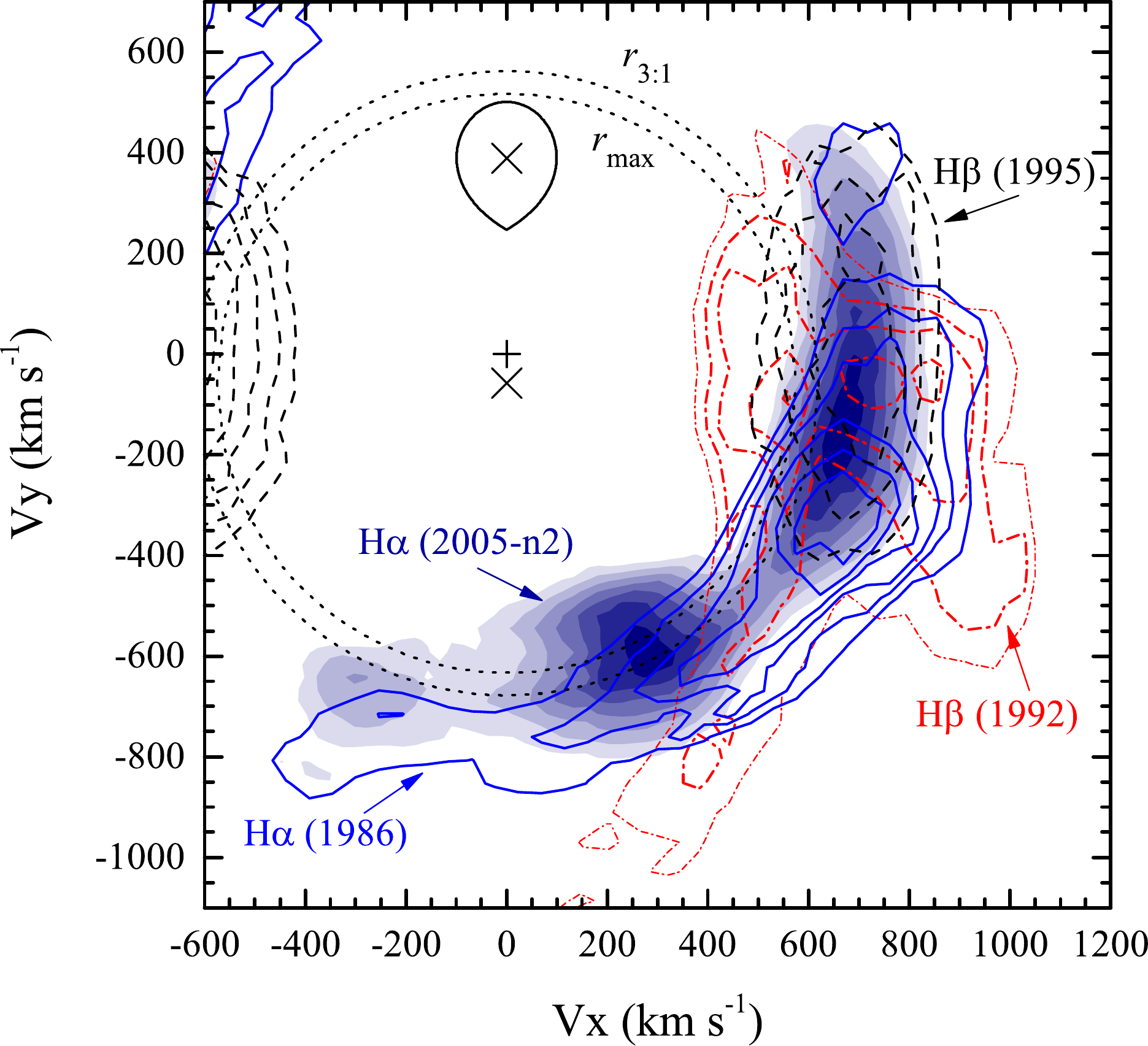}
}
\caption{Doppler maps that combine the tomograms for different lines from the 2005 observations (left)
         and for the strongest lines from different sets of observations (right). The maps are zoomed
         around the emission region in the leading side of the accretion disc. The circular dashed
         lines represent Keplerian velocities at the tidal truncation ($r_{\rm max}$) and 3:1 resonance
         ($r_{3:1}$) radii.
         }
\label{Fig:BrightSpot2}
\end{figure*}

\subsection{Spiral arm in the trailing side of the accretion disc}

The weakness of the hot spot in the \Halpha\ line makes it possible to trace the extended emission
structure in the fourth quadrant of the tomograms in detail. Figure~\ref{Fig:BrightSpot1} (right-hand
panel) shows a zoomed part of the \Halpha\ Doppler map from the set-2005-n2 that is centred around this
area. The latter can also be traced in other spectral lines and data sets, some of which are denoted in
Figure~\ref{Fig:BrightSpot1} (right-hand panel) by different lines. The location and shape of the
structure is nearly the same in all the
lines. It starts at the hotspot area and extends downstream in azimuth for $\sim$60\degr. Its width
gradually decreases until this `tail' has completely disappeared.

The origin of this emission is not clear. A similar structure was observed, e.g. in \object{WZ~Sge}
\citep{SpruitWZSge} and \object{U~Gem} \citep{UndaSanzana}. This kind of tail cannot be due to stream–disc
overflow, which might produce an excess of emission along the path of the stream \citep{KunzeSpeith}.
On the other hand, a tail is to be expected as a consequence of the post-impact hydrodynamics
of the gas stream and could be caused by material that has settled into Keplerian motion downstream from
the hotspot \citep[for a discussion see][and references therein]{SpruitWZSge}. However, despite an overall
similarity in the tails in HT~Cas and WZ~Sge, they  are significantly different. The tail
in WZ~Sge indeed shows circular Keplerian velocities along its trail, whereas the velocity in HT~Cas
increases from $\sim$760 to 880 \kms\ (Fig.~\ref{Fig:BrightSpot1}, right-hand panel). This corresponds
to the range of distances from the WD from 0.30$a$ to 0.22$a$ (Fig.~\ref{Fig:GeoMaps}). In fact, this
type of behaviour resembles the signature generated by spiral waves in the disc, rather than the hotspot
tail in WZ~Sge. However, the properties of spiral structure in the accretion discs have been
particularly well studied by both numerical simulations and observations \citep[and references
therein]{SteeghsStehle}, and no spiral waves or shocks are predicted in quiescent discs.

The observed spiral feature could be associated with a tidally thickened sector of the disc that is
elevated owing to tidal distortions and being irradiated by the WD or inner disc
\citep[for a discussion see][and Section~\ref{Sec:LeadingSpot} below]{Ogilvie2002,UndaSanzana},
although this type of thickening is more likely in the outer disc than at the observed position.
Another possible explanation for the spiral arm in HT~Cas may come from the hydrodynamical simulations by
\citet{Bisikalo98}. They showed that the gas stream may penetrate the outer disc regions, producing
a so-called hot line (an extended shock wave) that interacts and mixes with the disc, allowing
matter to be deposited at the inner disc regions.

\subsection{Emission region in the leading side of the accretion disc}
\label{Sec:LeadingSpot}

From Figs.~\ref{Fig:dopmaps2005} and \ref{Fig:dopmapsAll}, it may appear that the emission region in
the second quadrant of tomograms is bright in the \Halpha\ line only, much weaker in \Hbeta,\ and
almost undetectable in the \HeI\ lines. This impression is somewhat misleading because of the chosen contrast
and colour scaling to emphasise different components of Doppler maps. We estimated the contribution of
this emission source to the total flux of strongest emission lines from the set-2005 (\Halpha, \Hbeta,\
and \HeI\ 5876) and set-1986 (\Halpha\ and \Hbeta)  and found nearly the same value of 5--7 per cent for
all of them. The region trails along the accretion disc ring for some 150\degr\ in azimuth and perhaps
has a multicomponent structure. The highest quality and resolution Doppler map of \Halpha\
from the set-2005-n2 (Fig.~\ref{Fig:dopmaps3D}) clearly shows  two distinct spots of similar brightness
-- in the bottom and lower-right sides of the disc ring -- while in some of the other tomograms one of these
two spots prevails over the other.

We compared the location of these structures to those found in different lines from the 2005 observations
and from different sets of observations. Figure~\ref{Fig:BrightSpot2} shows that at least the lower-right
spot is always observed in the same position (there is a weak sign of vertical shift for the bottom spot),
even though the accretion disc properties differ considerably from one set of observations to another
(see Section~\ref{Sec:Decrement}). This resembles the behaviour of another short-period CV, namely BZ~UMa,
in which the similar spot in the leading side of the accretion disc remained present at the same position
during all stages of the outburst, from quiescence to the maximum \citep{NeustroevBZ}.
The velocities of the emission region suggest its
origin in the outer accretion disc. The spatial map of HT~Cas indicates a relatively sharp inner border
of the region at $R_{bs,in}$$\approx$0.25$a$, whereas its outer parts can be traced up to the Roche lobe
radius \emph{exceeding} $r_{\rm max}$ (Figure~\ref{Fig:GeoMaps}). Though the latter property may well be an
artefact of reconstruction because of the finite spectral resolution, there is no doubt that the emission
comes from the outermost parts of the disc.

Being located at the opposite side of the accretion
disc, neither of the bright spots can be associated with
the interaction between the gas stream and the disc. No shock waves that can produce an excess of emission
in the bottom-right side of tomograms are predicted by hydrodynamical simulations, and none of the theories
foresee stable shocks in the quiescent accretion disc.
We propose that the leading side bright spots are caused by irradiation of relatively compact thickened
sectors of the outer disc by the WD and/or hot, inner disc regions. The reason for this thickening is
not clear, but can be assumed to be tidally induced. A clue to understanding the exact process can
perhaps be gained from the fact that the outer parts of a large accretion disc are under the gravitational
influence of the secondary star. This prevents the disc from growing above the tidal radius $r_{\rm max}$,
where the tidal and viscous stresses are comparable (\citealt{Warner:1995aa}). How that truncation occurs
and how the disk thickness varies along the outer edge has not yet been well established. To the best of our
knowledge, no detailed 3D numerical simulations have been devoted to these questions so far. However,
\citet{Bisikalo98} pointed out an important role of the circum-disc halo that is created by matter which went
outside $r_{\rm max}$ and left the accretion disc. In their hydrodynamical simulations, the accretion disc
has a quasi-elliptical shape that extended in a direction opposite to the hot spot (see also \citealt{Kononov}).
\citet{Truss} also reported the appearance of similarly oriented elliptical discs, although only in extreme
mass-ratio compact binaries ($q$$<$0.1).

The asymmetry of the disc can also explain the shift of the radial velocity curve relative to the inferior
conjunction of the secondary star, as detected in HT~Cas (Section~\ref{Sec:RadVel}) and many other short-period
CVs \citep{WZ2}.
The fact that the shift is observed in a wide range of distances from the WD suggests significantly
asymmetric structures exist, even in the inner parts of the disc.

\subsection{Large accretion discs in cataclysmic variables}

The discovery of the extended bright area in the leading side of the accretion disc of HT~Cas enriches
the list of objects in which this type of feature has been observed. Based on the large measured radius of the disc
in HT~Cas, we make a guess in the previous subsection that the leading side bright spot was caused by
tidally induced thickened sectors of the outer large disc. In this respect it might be worth examining
if other systems that show a similar emission feature also have the large accretion disc.

We inspected several CVs with relatively well measured system parameters from the list given in the Introduction,
and found that most of them support our idea. For example, the disc in \object{VW~Hyi} appears slightly
larger than the tidal radius $r_{\rm max}$ \citep{VW_Hyi}, whereas in \object{WZ~Sge}, \object{V406~Vir},
\object{EZ~Lyn,} and the old nova \object{RR~Pic,} the disc radius was shown to be no smaller than the 3:1
resonance radius \citep{WZ1,SDSS1238,SDSS0804,RR_Pic2}. The disc radius of \object{IP~Peg}, as estimated
from the double-peaked emission line profiles ($\sim$570 \kms -- \citealt{NeustroevIP}), also appears to
be a bit larger than the tidal truncation radius (0.45$\pm0.05a$ and 0.40$a$, respectively).

Thus, most of the CVs with the leading side bright spot have the accretion disc, the radius of which is close to
the tidal truncation limit. It is worth noting that, at least in a few of them, the radius changes little
with time. For example, \citet{WZ2} claim that the same accretion disc radius has been observed in
WZ~Sge for 40 years. Our data also demonstrate that the radius of the disc in HT~Cas has not changed
much during all our observations, remaining consistently large. This contradicts  the modern understanding
of the evolution of the accretion disc through an outburst cycle, according to which the disc expands
during the outburst and then contracts with time \citep{Warner:1995aa}. The conclusion, that the disc
radius in many short-period CVs is close or even larger than the 3:1 resonance radius, has important
implications that relate to the observational properties of such systems.
It is expected that, when the
accretion disc expands beyond the 3:1 resonance radius, this causes the disc to become quasi-elliptical
and precess. The enhanced tidally-driven viscous dissipation in the disc, varying on the beat between
the orbital and disc precession periods, can result in superhumps in the light-curve \citep{Osaki}.
Thus, if it occurs during the quiescent state, then quiescent superhumps can be present. However, the
detectability of such superhumps should very much depend on the physical conditions in the outer disc.
For instance, it is difficult to expect strong quiescent superhumps in the lightcurve of HT~Cas, whose
outer accretion disc contributes a negligible amount of the total broadband light.

\section{Summary}

We have presented multi-epoch, time-resolved optical spectroscopic observations of the dwarf nova HT~Cas,
obtained during 1986, 1992, 1995, and 2005, with the aim of studying the properties of emission structures
in the system. Though HT~Cas has always remained in quiescence, its mean brightness has changed substantially
between the observations. The spectra are dominated by very strong and broad double-peaked emission
lines of the Balmer series. Numerous weaker lines of neutral helium and singly ionized iron (\ion{Fe}{II})
are also present. The high-excitation line of \HeII\ \l4686 is clearly detected.
A comparison of the averaged spectra from different data sets shows significant quantitative differences
between them. There are notable variations in both emission-line strengths and their ratios for different
lines, which indicates a variable mix of optically thin and thick conditions. Nevertheless, these variations
do not correlate with the system flux.

The emission lines are very broad with a FWZI of up to 5--6 thousand \kms\ and a peak-to-peak separation
of $\gtrsim$ 1100 \kms. We determined that the accretion disc radius, measured from the double-peaked
profiles, is consistently large and lies within the range of 0.45$-$0.52$a$. This is close to the tidal
truncation radius $r_{\rm max}$=0.52$a$ and slightly larger than the 3:1 resonance radius $r_{\rm 3:1}$
of 0.45$a$. This result is not consistent with previous radius measurements.

The radial velocity semi-amplitude of the WD was found to be $K_1=61\pm8$ \kms\ from the motion of the
wings of the emission lines. This value is very  consistent with that predicted by \citet{Horne91}.
However, the radial velocity curves of all the investigated emission lines are significantly shifted
relative to the eclipse ($\sim$55\degr), and therefore these lines cannot be used to represent the motion
of the WD. The shift is observed from the profile peaks to the extreme wings, suggesting a global asymmetry
may exist in the accretion disc of HT~Cas.

An extensive set of Doppler maps has revealed a very complex emission structure of the accretion disc.
Apart from a ring of disc emission, the tomograms display at least three areas of enhanced emission:
the hotspot from the area of interaction between the gas stream and the accretion disc, which is
superposed onto the elongated spiral structure, and the extended bright region on the leading side of
the disc, opposite  the location of the hotspot.

The position of the hotspot in all the emission lines is consistent with the trajectory of the gas stream.
However, the peaks of emission are located in the range of distances $R_{hs}$$\approx$0.22--0.30$a$, which
are much closer to the WD  than the disc edge (0.52$a$). This suggests that the outer disc regions have a
very low density, allowing the gas stream to flow almost freely before it starts to be seen as an emission
source. The  spiral arm appears  as a consequence of such a penetration. The stream produces an extended
shock wave that interacts and mixes with the disc.

The extended emission region in the leading side of the disc has been observed in many CVs, but it has had no
plausible explanation until now. We  found that in all the emission lines of HT~Cas, this structure is
always observed in the same position -- at the very edge of the large disc. Observations of other CVs,
which show a similar emission structure in their Doppler maps, seem to confirm this conclusion. We
propose that the leading side bright region is caused by irradiation of tidally thickened sectors of
the outer disc by the WD and/or hot inner disc regions.

\begin{acknowledgements}
The authors would like to thank Valery Suleimanov for  useful comments and Natalia Neustroeva for
help in preparing the manuscript.
We are thankful to the anonymous referee for their careful reading of the manuscript.
This work was supported by PAPIIT grants IN-100614 and CONACyT grants 151858, and CAR 208512 for
resources provided for this research.
\end{acknowledgements}

\bibliographystyle{aa} 
\bibliography{ht.bib} 
\end{document}